\documentclass[twocolumn,aps,showpacs,prl,superscriptaddress]{revtex4-1}
\usepackage[T1]{fontenc}
\usepackage[latin9]{inputenc}
\usepackage{amsmath}
\usepackage{amssymb}
\usepackage{esint}
\usepackage{graphicx}
\usepackage[normalem]{ulem}
\PassOptionsToPackage{normalem}{ulem}
\usepackage{pdfpages}
\usepackage{times}

\makeatletter

\usepackage{color}

\makeatother

\begin{document}

\title{Finite-temperature hydrodynamics for one-dimensional Bose gases: \\Breathing
mode oscillations as a case study}

\author{I. Bouchoule}

\affiliation{Laboratoire Charles Fabry, Institut d'Optique, CNRS, Univesité Paris
Sud 11, 2 Avenue Augustin Fresnel, F-91127 Palaiseau Cedex, France}

\author{S. S. Szigeti}

\affiliation{University of Queensland, School of Mathematics and Physics,
Brisbane, Queensland 4072, Australia}
\affiliation{ARC Centre of Excellence for Engineered Quantum Systems, University of Queensland, Brisbane, Queensland 4072, Australia}

\author{M.~J.~Davis}

\affiliation{University of Queensland, School of Mathematics and Physics,
Brisbane, Queensland 4072, Australia}

\author{K.~V.~Kheruntsyan}

\affiliation{University of Queensland, School of Mathematics and Physics,
Brisbane, Queensland 4072, Australia}
\begin{abstract}
We develop a finite-temperature hydrodynamic approach for a harmonically trapped one-dimensional quasicondensate and apply it to describe the phenomenon of frequency doubling in the breathing-mode oscillations of the quasicondensate momentum distribution. The doubling here refers to the oscillation frequency relative to the oscillations of the real-space density distribution, invoked by a sudden confinement quench. 
By constructing a nonequilibrium phase diagram that characterises the regime of frequency doubling and its gradual disappearance, we find that this crossover is governed by the quench strength and the initial temperature, rather than by the equilibrium-state crossover from the quasicondensate to the ideal Bose gas regime. The hydrodynamic predictions are supported by the results of numerical simulations based on a finite-temperature $c$-field approach, and extend the utility of the hydrodynamic theory for low-dimensional quantum gases to the description of finite-temperature systems and their dynamics in momentum space.
\end{abstract}


\date{\today}

\maketitle
Hydrodynamics is a powerful and broadly applicable
approach for characterizing the collective nonequilibrium behavior
of a wide range of classical and quantum fluids, including Fermi liquids, liquid helium, and ultra-cold atomic Bose and Fermi gases
\cite{LandauLifshitz,Pitaevskii-Stringary-book,Griffin-yellow-book,Griffin:1997,Kagan:1997,Pitaevskii:1998}. 
For ultra-cold gases, the hydrodynamic approach
has been particularly successful in describing the breathing (monopole) and higher-order (multipole) collective oscillations 
of harmonically trapped three-dimensional (3D) 
Bose-Einstein condensates  \cite{Stringari:1996,Pitaevskii:1998,Pitaevskii-Stringary-book}. 
For condensates near zero temperature, 
the applicability of the approach 
stems from the fact that for long-wavelength (low-energy) excitations
the hydrodynamic equations are essentially equivalent to those of superfluid hydrodynamics, which in turn can be derived from the Gross-Pitaevskii equation for the order parameter. 
For partially condensed samples at finite temperatures, the hydrodynamic equations should be generalized to the equations of two-fluid hydrodynamics, where the applicability of the approach to the normal 
(thermal) component of the gas 
is justified by fast thermalization times due to collisional relaxation \cite{Griffin-yellow-book,sidorenkov_second_2013}.

In contrast to 3D systems, the applicability of the hydrodynamic approach to 1D Bose gases is not well established. Firstly, in the thermodynamic limit 1D Bose gases lack the long-range order required for superfluid hydrodynamics to be \emph{a priori} applicable.  Secondly, the very notion of local thermalisation, required for the validity of collisional hydrodynamics of normal fluids, is questionable due to the underlying integrability of the uniform 1D Bose gas model \cite{Lieb:1963}. 
Despite these reservations, the hydrodynamic approach has already been 
applied to zero-temperature ($T=0$) dynamics of 1D Bose gases in
various scenarios \cite{Minguzzi-hydro-2001,Menotti:2002,
  Santos:2002,*Santos:2003, Peotta:2014, Choi:2015, Campbell:2015} 
(for related experiments, see
\cite{Moritz:2003,Naagerl:2009,Fang:2014}). 
The comparison of  hydrodynamic predictions with exact theoretical results is challenging.
In Ref. \cite{Peotta:2014}, time-dependent density matrix renormalization group simulations of the collision of 1D Bose gases at $T = 0$ found reasonable agreement with the hydrodynamic approximation, although the latter failed to predict short wavelength dynamics such as shock waves.
An alternative approximate  approach, based on the conservation of Lieb-Liniger rapidities, has been applied to describe the free expansion dynamics of a $T=0$ 1D gas~\cite{Campbell:2015} and was able to reproduce the hydrodynamic results for both weak and strong interactions.

At finite temperatures finding exact predictions is extremely difficult, and thus
developing a hydrodynamic approach is appealing, despite its lack of justification.
Here, we develop a general finite-$T$ hydrodynamic approach suitable for 1D Bose gases and specifically apply it to the breathing-mode oscillations of a harmonically trapped 1D quasicondensate. We find the predictions agree both with experimental observations \cite{Fang:2014} and numerical simulations of a finite-temperature $c$-field methodology \cite{Davis:2001b,Blakie:2008}. More remarkably, our hydrodynamic approach not only adequately describes the dynamics of the density distribution of the gas (the standard observable of the hydrodynamic theory), but it can be also used to describe the dynamics of the \emph{momentum distribution}. This is a key observable for quantum gas experiments, and has not previously been accessible from a hydrodynamic approach.

Reference~\cite{Fang:2014} experimentally studied confinement quenches
of a finite-$T$ 1D Bose gas. The key finding was the phenomenon of
frequency doubling in the oscillations of the momentum distribution
relative to the breathing-mode oscillations of the real-space density
profile.  For the experimental dataset deep in the quasicondensate regime,
a periodic narrowing of the momentum distribution occurred at twice
the frequency of the breathing mode of the density profile.  
Although finite-temperature effects are crucial for understanding 
the momentum-space properties of \textit{equilibrium} quasicondensates 
\cite{Petrov:2000,Richard:2003,Amerongen:2008,Hofferberth:2007,Jacqmin:2011,Armijo:2011}, the 
said experimental data for \textit{dynamics} were well-described
 by a simple zero-temperature classical hydrodynamic approach, wherein
 the frequency doubling was interpreted as a result of a self-reflection mechanism due to the 
mean-field interaction energy barrier. In  contrast to this behavior, no frequency doubling was observed in the nearly ideal
  Bose gas regime, as expected for a noninteracting gas. The experimentally observed 
  smooth crossover from the regime of frequency doubling to no doubling has so far not been explained theoretically.
  
Here, we explain this phenomenon within the hydrodynamic approach and construct, for the first time, the corresponding nonequilibrium 
  phase diagram, showing that the frequency doubling crossover is governed by 
  the quench strength and a nontrivial combination of the temperature and interaction strength. For small enough quenches, the crossover from frequency doubling to no doubling
can lie entirely within the quasicondensate regime, and does not require an equilibrium-state crossover to the ideal Bose gas regime. Constructing and studying phase diagrams is an important goal in many areas of physics, and our findings here serve as an example where equilibrium and dynamical phase diagrams are not identical.
We confirm our predictions by comparing the hydrodynamic results to those obtained numerically using finite-temperature $c$-field  
simulations (for a review, see \cite{Blakie:2008}) based on the projected Gross-Pitaevskii equation (PGPE) \cite{Davis:2001b}.

\emph{1. Hydrodynamic equations and evolution of the density distribution.}---The hydrodynamic 
approach relies on the local density approximation (LDA) and assumes
that the 1D system can be divided into small locally uniform slices, each of which is in  thermal equilibrium in 
the local moving frame. Moreover, one can assume that heat transfer between the slices is negligible for long-wavelength excitations
\cite{heat-transfer}, which
implies that each slice undergoes isentropic
(de)compression. The hydrodynamic description of this system is \cite{LandauLifshitz}
\begin{subequations} 
\begin{eqnarray}
\partial_{t}\rho+\partial_{x}(\rho v) & = & 0,\label{eq:HDEa}\\
\partial_{t}v+v\partial_{x}v & = & -\frac{1}{m}\partial_{x}V(x,t)-\frac{1}{m\rho}\partial_{x}P,\label{eq:HDEb}\\
\partial_{t}s+v\partial_{x}s & = & 0,\label{eq:HDEc}
\end{eqnarray}
\end{subequations} 
where $\rho(x,t)$ is the local 1D density of the slice at position $x$, $v(x,t)$
is the respective hydrodynamic velocity, $s(x,t)$ is the entropy per particle,
$P(x,t)$ is the pressure, 
$m$ is the mass of the constituent particles,
and $V(x,t)$ is the external trapping potential which for our case study is
harmonic, $V(x,t)=\frac{1}{2}m\omega(t)^{2}x^{2}$, of frequency $\omega(t)$.

We now apply the hydrodynamic
approach to 
describe the post-quench dynamics induced by  
the following scenario. Initially the atomic cloud with density profile $\rho_0(x)$ is 
in thermal equilibrium at temperature 
$T_0$ in the trap of frequency $\omega_0$.  Subsequently, at time $t\!=\!0$, the 
trap frequency is suddenly changed to $\omega_{1}$. 
To characterize the ensuing dynamics in different regimes of the 1D Bose gas, 
we introduce the dimensionless interaction parameter  $\gamma_{0}=mg/\hbar^2\rho_{0}(0)$
 and the dimensionless 
temperature 
$t_0=2\hbar^2k_B T_0/mg^2$ \cite{Kheruntsyan:2005,Bouchoule:2007}, where
$g$ is the coupling strength of the pairwise $\delta$-function interaction 
potential. The solutions of the HDEs (\ref{eq:HDEa})-(\ref{eq:HDEc}) describing this 
harmonic-confinement quench 
 depend only on the thermodynamic equation of state of the gas.
In each of the following three cases, (i) --  ideal gas regime 
($t_0,\gamma_0^{3/2}t_0\!\gg\! 1$), 
(ii) -- strongly interacting or Tonks-Girardeau regime ($\gamma_0,1/t_0\gg 1$), and 
(iii) -- quasicondensate regime ($\gamma_0,\gamma_0^{3/2}t_0\ll 1$),
the solutions  reduce to  scaling equations  of the form
\begin{gather}
\rho(x,t)=\rho_{0}(x/\lambda(t))/\lambda(t),\;\;\;\;v(x,t)=x\dot{\lambda}(t)/\lambda(t),\label{eq:scaling-rho-v}\\
T(t)=T_{0}/\lambda(t)^{\nu+1},\label{eq:scaling-T}
\end{gather}
where the scaling parameter $\lambda(t)$ [with $\dot{\lambda}\!\equiv\! d\lambda(t)/dt$, $\lambda(0)\!=\!1$, 
and $\dot{\lambda}(0)\!=\!0$] satisfies the ordinary differential equation,
\begin{equation}
\ddot{\lambda}=-\omega_{1}^{2}\lambda+\omega_{0}^{2}/\lambda^{2\nu+1},
\label{eq-for-lambda-general}
\end{equation}
with the value of $\nu$ in different regimes given below {\cite{footnote-arbitrary-omega}. The hydrodynamic solution~(\ref{eq:scaling-T}) for the temperature 
is one of the key results of this paper as it allows one to simply calculate the evolution of the temperature-dependent momentum distribution 
of the gas (see below).

(i) \emph{Ideal gas regime} ($t_0,\gamma_0^{3/2}t_0\!\gg\! 1$): In
this case $\nu=1$, and the validity of the above scaling solutions 
can be
demonstrated using a dimensional analysis  of the 
equation of state (see Ref. \cite{sup}), which we note is also applicable to an ideal Fermi gas. 
 Equation (\ref{eq-for-lambda-general}) in this regime has an
explicit analytic solution,
\begin{equation}
\lambda(t)=\sqrt{1+\epsilon \sin^2(\omega_1t )}.
\label{eq:lambda-IBG}
\end{equation}
This corresponds to harmonic oscillations of the mean squared width of the density profile, occurring
at frequency $\omega_{B}=2\omega_{1}$,
with $\epsilon\equiv(\omega_{0}/\omega_{1})^{2}-1$ characterizing the quench strength. 
This result coincides with that for a noninteracting gas obtained from a single-particle picture.
The fact that the hydrodynamic approach, which
{\it a priori} assumes sufficient collisions to ensure local  thermal equilibrium, 
agrees with the results for a noninteracting gas 
is specific to the harmonic-confinement quench 
considered here and is accidental.

(ii) \emph{Strongly interacting regime} ($\gamma_0,1/t_0\!\gg\! 1$): Here, 
the equation of state is that of an ideal Fermi gas so that the previous ideal gas results 
apply, and Eqs. (\ref{eq:scaling-rho-v})--(\ref{eq-for-lambda-general}) are fulfilled with $\nu=1$.
The breathing mode oscillations of
the momentum distribution of a finite-temperature Tonks-Girardeau gas are discussed elsewhere
\cite{Tonks-breathing-paper}.

(iii) \emph{Quasicondensate regime} ($\gamma_0,\gamma_0^{3/2}t_0\!\ll \!1$): In this case
$\nu\!=\!1/2$, and the validity of the scaling solutions
(\ref{eq:scaling-rho-v}) can be demonstrated using the 
equation of state $P\!=\!\frac{1}{2}g\rho^{2}$. The latter can be derived from
the quasicondensate chemical potential $\mu\!=\!g\rho$ and
the Gibbs-Duhem  relation $\rho\!=\!(\partial P/\partial
\mu)_T$. For a weak quench,
  $\epsilon \!\ll \!1$, the solution to Eq.~(\ref{eq-for-lambda-general})  oscillates at frequency
  $\omega_{B}\!\simeq\!\sqrt{3}\omega_{1}$ and is nearly harmonic with an
  amplitude $\lambda(t\!=\!\pi/\omega_B)-1 \!\simeq\!
  2\epsilon/3$ \cite{small-amplitude}.  According to Eq.~(\ref{eq:scaling-rho-v}), 
the density
profile  breathes self-similarly, maintaining
its initial  Thomas-Fermi 
parabolic shape, $\rho_{0}(x)=\rho_{0}(0)(1-x^{2}/X_{0}^{2})$ for $x\leq X_{0}$
{[}$\rho_{0}(x)=0$ otherwise{]}, with
$X_{0}\!=\!\sqrt{2g\rho_{0}(0)/m\omega_{0}^{2}}$.
Finite-temperature effects
are not seen in the dynamics of the density
distribution $\rho(x,t)$ \cite{qBEC-thermal-effects} because in this regime the equation of 
state 
does not depend on the temperature. 
However, as we show below, such effects can be revealed in the
dynamics of the momentum distribution.

\emph{2. Dynamics of the momentum distribution.}---
Let us consider a slice of the gas in the region
  $[x, x+dx]$ of density $\rho(x,t)$, velocity $v(x,t)$, and entropy
  per particle $s(x,t)$. In the laboratory frame its  momentum distribution is $\overline{n}(\rho, s, k - m v / \hbar)$,
  where $\overline{n}$ is the equilibrium momentum distribution of a
  homogeneous gas, which we normalize to $\int dk \,
  \overline{n}(\rho, s; k) = \rho$. The total momentum distribution
is then given by
\begin{equation}
	n(k,t)=\int dx\,\overline{n}(\rho,s;k-mv(x,t)/\hbar)\label{eq:nkt-gen}.
\end{equation}

There are two contributions to $n(k,t)$: the hydrodynamic velocity field and the 
contribution of thermal velocities, 
which have different effects on the breathing mode
oscillations.
In order to see the sole effect of the 
hydrodynamic velocity field,
 let us first disregard the 
effect of the thermal velocities, 
taking $\overline{n}(\rho,s;k-mv(x,t)/\hbar) = \rho \delta(k-mv(x,t)/\hbar)$, 
where $\delta(k)$ is the Dirac delta function.
If a scaling solution as in Eq.~(\ref{eq:scaling-rho-v}) exists,  
 the hydrodynamic component 
of the momentum distribution evolves according to
\begin{equation}
	n_{\mathrm{h}}(k,t) =  \frac{\hbar}{m|\dot{\lambda}|} \rho_0\!\left( \frac{\hbar k}{m \dot{\lambda}} \right).
	\label{nk-hydro}
\end{equation}
For oscillatory $\lambda(t)$, this implies that 
$n_{\mathrm{h}}(k,t)$ collapses to a zero-width distribution \emph{twice} per position-space density breathing cycle: when the width of the cloud in real space is both largest and smallest, both corresponding to $\dot{\lambda}=0$. Therefore, the oscillations of the hydrodynamic contribution to the momentum distribution will always display frequency doubling. 

Now consider
the additional contribution of thermal velocities to $n(k,t)$,
which changes as each slice undergoes isentropic compression and decompression 
during the breathing cycle.
Since one expects the momentum width to be a monotonically increasing function 
of the compression factor, 
the thermal momentum width of each slice [and hence of the overall momentum distribution $n(k,t)$] is expected to oscillate 
out-of-phase relative to the width of the real-space density profile, but at the same breathing 
frequency $\omega_B$.

The evolution of the overall momentum distribution $n(k,t)$ results from the combination of 
the hydrodynamic and thermal parts.  For a near-ideal gas at $T>0$ this  
  leads to a somewhat fortuitous cancellation of the hydrodynamic velocity field by the thermal component, and
  so  the total momentum distribution always oscillates at $\omega_B=2\omega_{1}$ (see Ref. \cite{sup}) and  never displays frequency doubling,  consistent with the single-particle picture.

The situation is different, however, in the quasicondensate regime.
The momentum distribution of  a homogeneous quasicondensate
of density $\rho$ and temperature $T$,
for wavelengths in the phononic regime (i.e., $k\!\ll \!\hbar/\sqrt{mg\rho}$)
is given by a Lorentzian $\overline{n}(\rho,s;k)\!=\!(2\rho l_{\phi}/\pi)/\left[1+(2l_{\phi}k)^{2}\right]$ \cite{sup}.
Substituting this Lorentzian into Eq.~(\ref{eq:nkt-gen}), we obtain the full momentum distribution 
of the trapped gas,
\begin{equation}
n(k,t)=\frac{1}{\pi}\int dx\,\frac{2l_{\phi}(x,t)\rho(x,t)}{1+4[l_{\phi}(x,t)]^{2}[k-mv(x,t)/\hbar]^{2}},
\label{eq:momentum-distr-qBEC}
\end{equation}
where $l_{\phi}(x,t)=\hbar^{2}\rho(x,t)/mk_{B}T(t)$.
According to the scaling solutions (\ref{eq:scaling-rho-v}) and 
(\ref{eq:scaling-T}) with $\nu=1/2$ (see also \cite{sup}),
$l_{\phi}(x,t)$
evolves as
\begin{equation}
l_{\phi}(x,t)=\sqrt{\lambda}\,l_{\phi}^{(0)}\tilde{\rho}_{0}(x/\lambda),\label{eq:scaling-lphi}
\end{equation}
where $\tilde{\rho}_{0}(x)=1-x^{2}/X_{0}^{2}$ is the scaled initial
density profile and $l_{\phi}^{(0)}=\hbar^{2}\rho_{0}(0)/mk_{B}T_{0}=2[\rho_{0}(0)\gamma_{0}^{2}t_{0}]^{-1}$.

Combining the scaling solution for $l_{\phi}(x,t)$ with that for $\rho(x,t)$, and changing variables to $u=x/\lambda X_{0}$
in Eq.~(\ref{eq:momentum-distr-qBEC}), leads to the following final
result 
\begin{equation}
n(k,t)\!=\! B
\sqrt{\tilde{\lambda}}\!\int_{-1}^{1}\!du\,\frac{(1-u^{2})^{2}}{1+4\tilde{\lambda}(1-u^{2})^{2}\left (\tilde{k}-\frac{\omega_{1}}{\omega_0}A\dot{\tilde{\lambda}}u\right )^{2}}.
\label{eq.nktqbec}
\end{equation}
Here,
$\tilde{k}\!=\!l_{\phi}^{(0)}k$, 
 $A\!=\!m\omega_0 X_{0}l_{\phi}^{(0)}/\hbar\!=\!\sqrt{8}/\gamma_0^{3/2} t_0 $,
and $B\!=\!2\rho_{0}(0)l_{\phi}^{(0)}X_{0}/\pi$ is 
a normalization factor. 
In addition, we have introduced a dimensionless time $\tau\equiv \omega_{1} t$, so that the dimensionless functions 
$\tilde{\lambda}(\tau)\!\equiv \!\lambda(\tau/\omega_{1})$ and 
$\dot{\tilde{\lambda}}= d\tilde{\lambda}/d\tau$, obtained from 
Eq.~(\ref{eq-for-lambda-general}),
depend only on the ratio $\omega_{1}/\omega_0$, or equivalently only on the quench strength $\epsilon \!= \!(\omega_0/\omega_{1})^2 - 1$.
Thus, for a given $\epsilon$, the
evolution of $n(k,t)$ is governed solely by the dimensionless parameter
$A$, which itself depends only on the initial intensive parameters $\gamma_0$ and $t_0$.
Note that $A\gg 1$ in the quasicondensate regime where
$\gamma_0^{3/2}t_0\!\ll\! 1$~\cite{Kheruntsyan:2005,Jacqmin:2011}.

Using Eq.~(\ref{eq.nktqbec}) for a given $A$ and 
quench strength $\epsilon$,
we can now compute the
evolution of the full momentum distribution and its half width at half
maximum (HWHM); see Figs.~\ref{fig.Afd}(a)--(b).  The HWHM can then be
fitted with a sum of two sinusoidal functions: the fundamental mode
oscillating at $\omega_B$ ($\simeq \sqrt{3}\omega_1$, for $\epsilon \ll 1$) and the first harmonic oscillating at
$2\omega_B$, with amplitudes $c_1$ and $c_2$, respectively. Defining
the weight of the fundamental mode as $K\!=\!c_1^2/(c_1^2+c_2^2)$, we
identify the frequency doubling phenomenon with $K\!\ll \!1 $,
whereas $K\simeq 1$ corresponds to the absence of doubling. 
The doubling crossover can,
therefore, be defined as the value of $A=A_{\mathrm{cr}}$ for which
$K=1/2$.
As we show in \cite{sup}, for small quench amplitudes 
one expects the frequency doubling to occur for $A\sqrt{\epsilon} \gg 1$, 
while for  $A\sqrt{\epsilon} \ll 1$  the thermal effects 
dominate and the frequency doubling is absent; accordingly, $A_{\mathrm{cr}}$ is expected to scale as 
$A_{\mathrm{cr}}\propto 1/\sqrt{\epsilon}$. 
Figure ~\ref{fig.Afd}(c) shows the nonequilibrium phase diagram of the crossover from frequency doubling to no doubling 
and confirms that $A_{\mathrm{cr}}$, obtained using Eq.~(\ref{eq.nktqbec}) and the fitting procedure described above, does indeed scale as $\propto 1/\sqrt{\epsilon}$.

\begin{figure*}[tbph]
\includegraphics[width=16.5cm]{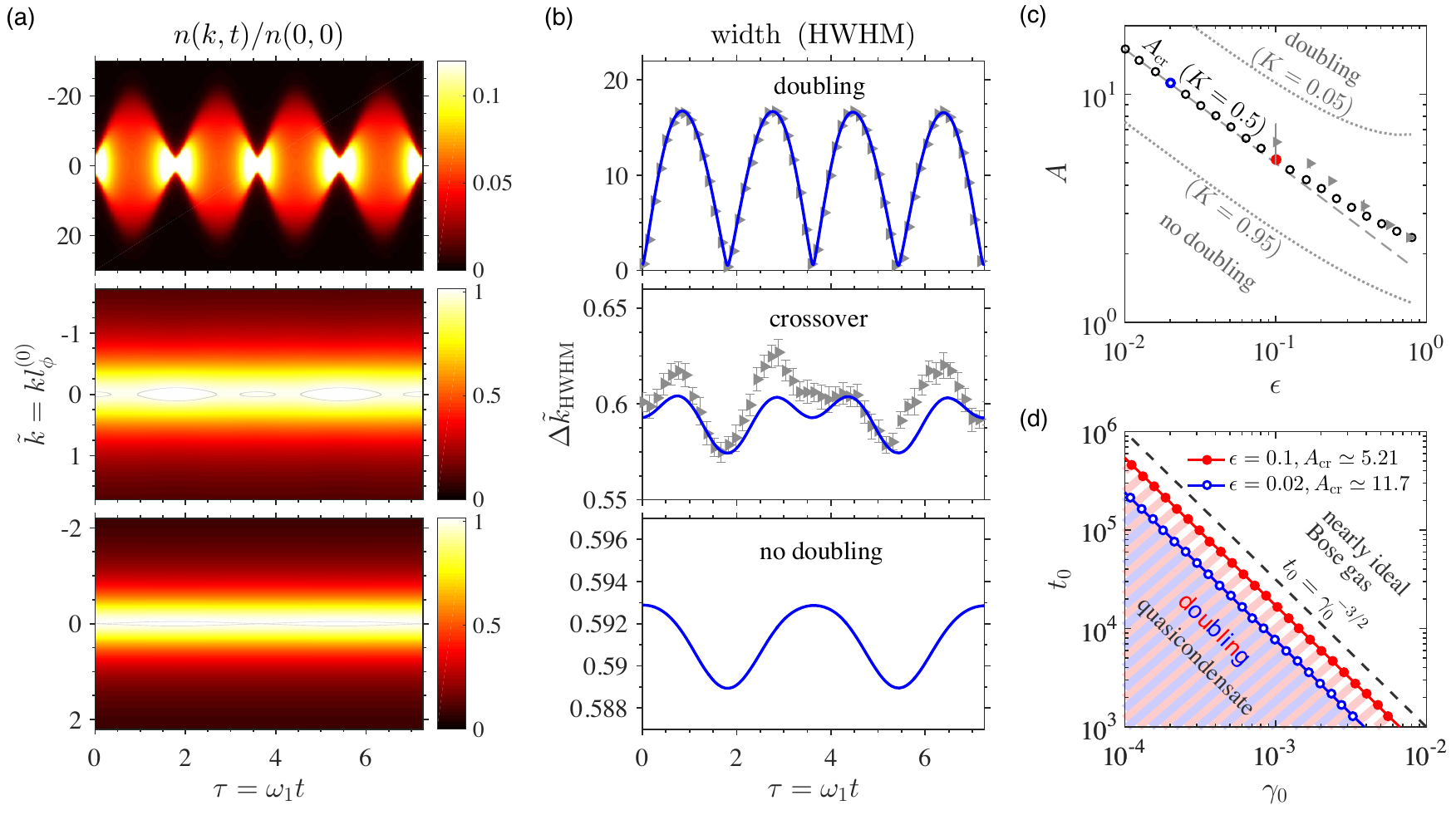}
\caption{(Color online). (a) Breathing mode oscillations and (b) the respective widths (HWHM) of the momentum distribution of a harmonically trapped 1D quasicondensate after a confinement quench as a function of the dimensionless time $\tau=\omega_{1} t$. The three examples shown in (a) and (b) correspond, respectively, to: $\epsilon=0.563$, $A=104$ (with $t_0 = 10^6$ and  $\hbar \omega_0 / [g \rho_0(0)] = 3.0 \times 10^{-3}$ is $c$-field simulations) -- top row; $\epsilon=0.101$, $A=A_{\mathrm{cr}}=5.21$ ($t_0 = 10^3$, $\hbar \omega_0 / [g \rho_0(0)] = 1.1 \times 10^{-2}$) -- middle row; and $\epsilon=0.0203$, $A=3.95$ -- bottom row \cite{c-field-c}. The grey triangles are the $c$-field data \cite{sup}, with the error bars indicating $95$\% confidence interval.
(c) Nonequilibrium phase diagram of the dynamical crossover from frequency doubling to no doubling in the $A$-$\epsilon$ parameter space; data points (circles) show the 
crossover values $A_{\mathrm{cr}}$
for which the weights of the fundamental and the first harmonics 
are equal ($K=1/2$); grey triangles are from $c$-field simulations.  The dashed line is a fit in the region $A>5$ with a power law $A_{\mathrm{cr}}\simeq 1.58 /\sqrt{\epsilon}$ (see text), whereas the two dotted lines show the values of $A$ corresponding to $K=0.05$ and $K=0.95$. (d) Frequency doubling conditions 
superimposed on the \textit{equilibrium} phase diagram of the 1D Bose gas \cite{Kheruntsyan:2005,Jacqmin:2011}, drawn in terms of the dimensionless temperature ($t_0$) and interaction strength ($\gamma_0$), and covering the quasicondensate and the neighbouring nearly ideal Bose gas regimes. The grey dashed line ($t_0=\gamma_0^{-3/2}$) corresponds to the crossover between the two equilibrium regimes. The two lines with filled (red) and open (blue) circles, on the other hand, correspond to the frequency doubling crossover conditions for two different quenches, $\epsilon=0.1$ and $\epsilon=0.02$ (the respective data points in (c) are labelled in the same way). The (light red and light blue) shaded areas underneath these lines correspond to the conditions where the frequency doubling occurs.}
\label{fig.Afd}
\end{figure*}

In Fig.~\ref{fig.Afd}(d) we superimpose the
conditions for observing frequency doubling on
the equilibrium phase diagram of the 1D Bose gas.
As we see, for small enough quench strengths, the crossover from doubling to 
no doubling lies well within the quasicondensate regime. 
We therefore conclude that this phenomenon is governed not by 
 the crossover from the ideal Bose gas regime into the quasicondensate regime,
but by the
competition between the hydrodynamic velocity (which
always displays doubling) and the broadening/narrowing
of the thermal component of the gas due to adiabatic
compression/decompression (which always oscillates at the fundamental
frequency $\omega_{B}$).

Although the applicability of the hydrodynamic theory in this system
might be questionable, our analytic results have been benchmarked
against finite-temperature $c$-field simulations, 
whose validity for 
 degenerate weakly interacting Bose gases is well established
\cite{Davis:2001b,Blakie:2008,Davis:2012,sinatra2000monte,Castin:2000}.
In this approach,
  the Bose gas is approximated as a classical field, whose evolution is
   governed by the time-dependent Gross-Pitaevskii equation 
(GPE), with the initial state being sampled from the classical Gibbs ensemble 
for the given temperature and density~\cite{sup}.
Qualitatively, the same behaviour as in
Figs.~\ref{fig.Afd}(a)--(b) based on the hydrodynamic approach 
occurs in our $c$-field simulations
\cite{c-field-c}; quantitatively, the crossover from doubling to
no-doubling is in broad agreement with the analytic predictions [see
  Fig.~\ref{fig.Afd}(c)]. Moreover, as we argue in Ref.~\cite{sup},
for sufficiently weak confinement (small $\omega_0$), the $c$-field
dynamics are governed by just two dimensionless parameters, $A$ and
$\epsilon$, as predicted from the hydrodynamic approach. Overall, the
performance of the hydrodynamic theory---as validated by our $c$-field simulations---in modelling the harmonic
confinement quench of a finite-temperature quasicondensate is
remarkable. Moreover, even though the hydrodynamic results 
of Eq.~(\ref{eq.nktqbec}) formally require $A\!\gg \!1$ to ensure the applicability of the 
quasicondensate regime, 
our comparison with 
$c$-field simulations shows that Eq.~(\ref{eq.nktqbec}) continues to give accurate predictions even for moderate values of $A\gtrsim 1$.

In summary, we have developed a finite-temperature hydrodynamic approach for a harmonically trapped 1D Bose gas and applied 
it to the study of breathing mode oscillations in the quasicondensate regime. While the usual scope of the hydrodynamic theory is to
 describe the evolution of the real-space density of a gas, our approach extends its utility to describe the evolution of its momentum distribution.
The approach allowed us to discern the contribution of the hydrodynamic velocity field and that of thermal excitations in the oscillatory dynamics of the momentum distribution of the 1D quasicondensate, hence explaining the full mechanism behind the phenomenon of frequency doubling and the crossover to no doubling. The hydrodynamic predictions are in broad agreement with numerical simulations based on finite-temperature $c$-field simulations. Our approach can address not only the sudden quench scenario studied here, but also the dynamics under arbitrary driving of the trapping frequency $\omega(t)$, in which case the differential equation for the scaling parameter $\lambda(t)$, Eq.~(\ref{eq-for-lambda-general}), must be solved numerically. Future extensions of this work will concern the treatment of breathing mode oscillations in the strongly interacting regime \cite{Tonks-breathing-paper}, and could also address collective behavior of 1D Bose gases in anharmonic traps, as well as of 2D and 3D quasicondensates in highly-elongated geometries.

K\;.V.\;K. acknowledges stimulating discussions with D.\;M. Gangardt.
I.\;B. acknowledges support by the Centre de Comp\'{e}tences Nanosciences \^{I}le-de-France; S.\;S.\;S. -- by the Australian Research Council
Centre of Excellence for Engineered Quantum Systems (grant No.~CE110001013); M.\;J.\;D. and K.\;V.\;K. -- by the ARC Discovery
Project grants DP160103311 and DP140101763. Numerical simulations were performed using XMDS2 \cite{Dennis:2012} on the University of Queensland School of Mathematics and Physics computer ``Obelix'', with thanks to I. Mortimer for computing support.


\begin{thebibliography}{40}%
\makeatletter
\providecommand \@ifxundefined [1]{%
 \@ifx{#1\undefined}
}%
\providecommand \@ifnum [1]{%
 \ifnum #1\expandafter \@firstoftwo
 \else \expandafter \@secondoftwo
 \fi
}%
\providecommand \@ifx [1]{%
 \ifx #1\expandafter \@firstoftwo
 \else \expandafter \@secondoftwo
 \fi
}%
\providecommand \natexlab [1]{#1}%
\providecommand \enquote  [1]{``#1''}%
\providecommand \bibnamefont  [1]{#1}%
\providecommand \bibfnamefont [1]{#1}%
\providecommand \citenamefont [1]{#1}%
\providecommand \href@noop [0]{\@secondoftwo}%
\providecommand \href [0]{\begingroup \@sanitize@url \@href}%
\providecommand \@href[1]{\@@startlink{#1}\@@href}%
\providecommand \@@href[1]{\endgroup#1\@@endlink}%
\providecommand \@sanitize@url [0]{\catcode `\\12\catcode `\$12\catcode
  `\&12\catcode `\#12\catcode `\^12\catcode `\_12\catcode `\%12\relax}%
\providecommand \@@startlink[1]{}%
\providecommand \@@endlink[0]{}%
\providecommand \url  [0]{\begingroup\@sanitize@url \@url }%
\providecommand \@url [1]{\endgroup\@href {#1}{\urlprefix }}%
\providecommand \urlprefix  [0]{URL }%
\providecommand \Eprint [0]{\href }%
\providecommand \doibase [0]{http://dx.doi.org/}%
\providecommand \selectlanguage [0]{\@gobble}%
\providecommand \bibinfo  [0]{\@secondoftwo}%
\providecommand \bibfield  [0]{\@secondoftwo}%
\providecommand \translation [1]{[#1]}%
\providecommand \BibitemOpen [0]{}%
\providecommand \bibitemStop [0]{}%
\providecommand \bibitemNoStop [0]{.\EOS\space}%
\providecommand \EOS [0]{\spacefactor3000\relax}%
\providecommand \BibitemShut  [1]{\csname bibitem#1\endcsname}%
\let\auto@bib@innerbib\@empty
\bibitem [{\citenamefont {Landau}\ and\ \citenamefont
  {Lifshitz}()}]{LandauLifshitz}%
  \BibitemOpen
  \bibfield  {author} {\bibinfo {author} {\bibfnamefont {L.~D.}\ \bibnamefont
  {Landau}}\ and\ \bibinfo {author} {\bibfnamefont {E.~M.}\ \bibnamefont
  {Lifshitz}},\ }\href@noop {} {\emph {\bibinfo {title} {Fluid Mechanics}}}\
  (\bibinfo  {publisher} {2nd edition, Vol. 6 of A Course of Theoretical
  Physics, Pergamon Press, Oxford, 1987})\BibitemShut {NoStop}%
\bibitem [{\citenamefont {Pitaevskii}\ and\ \citenamefont
  {Stringari}()}]{Pitaevskii-Stringary-book}%
  \BibitemOpen
  \bibfield  {author} {\bibinfo {author} {\bibfnamefont {L.~P.}\ \bibnamefont
  {Pitaevskii}}\ and\ \bibinfo {author} {\bibfnamefont {S.}~\bibnamefont
  {Stringari}},\ }\href@noop {} {\emph {\bibinfo {title} {Bose-Einstein
  Condensation}}}\ (\bibinfo  {publisher} {Clarendon Press, Oxford,
  2003})\BibitemShut {NoStop}%
\bibitem [{\citenamefont {Griffin}\ \emph {et~al.}()\citenamefont {Griffin},
  \citenamefont {Nikuni},\ and\ \citenamefont {Zaremba}}]{Griffin-yellow-book}%
  \BibitemOpen
  \bibfield  {author} {\bibinfo {author} {\bibfnamefont {A.}~\bibnamefont
  {Griffin}}, \bibinfo {author} {\bibfnamefont {T.}~\bibnamefont {Nikuni}}, \
  and\ \bibinfo {author} {\bibfnamefont {E.}~\bibnamefont {Zaremba}},\
  }\href@noop {} {\emph {\bibinfo {title} {Bose-Condensed Gases at Finite
  Temperatures}}}\ (\bibinfo  {publisher} {Cambridge University Press,
  Cambridge, 2009})\BibitemShut {NoStop}%
\bibitem [{\citenamefont {Griffin}\ \emph {et~al.}(1997)\citenamefont
  {Griffin}, \citenamefont {Wu},\ and\ \citenamefont
  {Stringari}}]{Griffin:1997}%
  \BibitemOpen
  \bibfield  {author} {\bibinfo {author} {\bibfnamefont {A.}~\bibnamefont
  {Griffin}}, \bibinfo {author} {\bibfnamefont {W.-C.}\ \bibnamefont {Wu}}, \
  and\ \bibinfo {author} {\bibfnamefont {S.}~\bibnamefont {Stringari}},\ }\href
  {\doibase 10.1103/PhysRevLett.78.1838} {\bibfield  {journal} {\bibinfo
  {journal} {Phys. Rev. Lett.}\ }\textbf {\bibinfo {volume} {78}},\ \bibinfo
  {pages} {1838} (\bibinfo {year} {1997})}\BibitemShut {NoStop}%
\bibitem [{\citenamefont {Kagan}\ \emph {et~al.}(1997)\citenamefont {Kagan},
  \citenamefont {Surkov},\ and\ \citenamefont {Shlyapnikov}}]{Kagan:1997}%
  \BibitemOpen
  \bibfield  {author} {\bibinfo {author} {\bibfnamefont {Y.}~\bibnamefont
  {Kagan}}, \bibinfo {author} {\bibfnamefont {E.~L.}\ \bibnamefont {Surkov}}, \
  and\ \bibinfo {author} {\bibfnamefont {G.~V.}\ \bibnamefont {Shlyapnikov}},\
  }\href {\doibase 10.1103/PhysRevA.55.R18} {\bibfield  {journal} {\bibinfo
  {journal} {Phys. Rev. A}\ }\textbf {\bibinfo {volume} {55}},\ \bibinfo
  {pages} {R18} (\bibinfo {year} {1997})}\BibitemShut {NoStop}%
\bibitem [{\citenamefont {Pitaevskii}\ and\ \citenamefont
  {Stringari}(1998)}]{Pitaevskii:1998}%
  \BibitemOpen
  \bibfield  {author} {\bibinfo {author} {\bibfnamefont {L.}~\bibnamefont
  {Pitaevskii}}\ and\ \bibinfo {author} {\bibfnamefont {S.}~\bibnamefont
  {Stringari}},\ }\href {\doibase 10.1103/PhysRevLett.81.4541} {\bibfield
  {journal} {\bibinfo  {journal} {Phys. Rev. Lett.}\ }\textbf {\bibinfo
  {volume} {81}},\ \bibinfo {pages} {4541} (\bibinfo {year}
  {1998})}\BibitemShut {NoStop}%
\bibitem [{\citenamefont {Stringari}(1996)}]{Stringari:1996}%
  \BibitemOpen
  \bibfield  {author} {\bibinfo {author} {\bibfnamefont {S.}~\bibnamefont
  {Stringari}},\ }\href {\doibase 10.1103/PhysRevLett.77.2360} {\bibfield
  {journal} {\bibinfo  {journal} {Phys. Rev. Lett.}\ }\textbf {\bibinfo
  {volume} {77}},\ \bibinfo {pages} {2360} (\bibinfo {year}
  {1996})}\BibitemShut {NoStop}%
\bibitem [{\citenamefont {Sidorenkov}\ \emph {et~al.}(2013)\citenamefont
  {Sidorenkov}, \citenamefont {Tey}, \citenamefont {Grimm}, \citenamefont
  {Hou}, \citenamefont {Pitaevskii},\ and\ \citenamefont
  {Stringari}}]{sidorenkov_second_2013}%
  \BibitemOpen
  \bibfield  {author} {\bibinfo {author} {\bibfnamefont {L.~A.}\ \bibnamefont
  {Sidorenkov}}, \bibinfo {author} {\bibfnamefont {M.~K.}\ \bibnamefont {Tey}},
  \bibinfo {author} {\bibfnamefont {R.}~\bibnamefont {Grimm}}, \bibinfo
  {author} {\bibfnamefont {Y.-H.}\ \bibnamefont {Hou}}, \bibinfo {author}
  {\bibfnamefont {L.}~\bibnamefont {Pitaevskii}}, \ and\ \bibinfo {author}
  {\bibfnamefont {S.}~\bibnamefont {Stringari}},\ }\href {\doibase
  10.1038/nature12136} {\bibfield  {journal} {\bibinfo  {journal} {Nature}\
  }\textbf {\bibinfo {volume} {498}},\ \bibinfo {pages} {78} (\bibinfo {year}
  {2013})}\BibitemShut {NoStop}%
\bibitem [{\citenamefont {Lieb}\ and\ \citenamefont
  {Liniger}(1963)}]{Lieb:1963}%
  \BibitemOpen
  \bibfield  {author} {\bibinfo {author} {\bibfnamefont {E.~H.}\ \bibnamefont
  {Lieb}}\ and\ \bibinfo {author} {\bibfnamefont {W.}~\bibnamefont {Liniger}},\
  }\href {\doibase 10.1103/PhysRev.130.1605} {\bibfield  {journal} {\bibinfo
  {journal} {Phys. Rev.}\ }\textbf {\bibinfo {volume} {130}},\ \bibinfo {pages}
  {1605} (\bibinfo {year} {1963})}\BibitemShut {NoStop}%
\bibitem [{\citenamefont {Minguzzi}\ \emph {et~al.}(2001)\citenamefont
  {Minguzzi}, \citenamefont {Vignolo}, \citenamefont {Chiofalo},\ and\
  \citenamefont {Tosi}}]{Minguzzi-hydro-2001}%
  \BibitemOpen
  \bibfield  {author} {\bibinfo {author} {\bibfnamefont {A.}~\bibnamefont
  {Minguzzi}}, \bibinfo {author} {\bibfnamefont {P.}~\bibnamefont {Vignolo}},
  \bibinfo {author} {\bibfnamefont {M.~L.}\ \bibnamefont {Chiofalo}}, \ and\
  \bibinfo {author} {\bibfnamefont {M.~P.}\ \bibnamefont {Tosi}},\ }\href
  {\doibase 10.1103/PhysRevA.64.033605} {\bibfield  {journal} {\bibinfo
  {journal} {Phys. Rev. A}\ }\textbf {\bibinfo {volume} {64}},\ \bibinfo
  {pages} {033605} (\bibinfo {year} {2001})}\BibitemShut {NoStop}%
\bibitem [{\citenamefont {Menotti}\ and\ \citenamefont
  {Stringari}(2002)}]{Menotti:2002}%
  \BibitemOpen
  \bibfield  {author} {\bibinfo {author} {\bibfnamefont {C.}~\bibnamefont
  {Menotti}}\ and\ \bibinfo {author} {\bibfnamefont {S.}~\bibnamefont
  {Stringari}},\ }\href {\doibase 10.1103/PhysRevA.66.043610} {\bibfield
  {journal} {\bibinfo  {journal} {Phys. Rev. A}\ }\textbf {\bibinfo {volume}
  {66}},\ \bibinfo {pages} {043610} (\bibinfo {year} {2002})}\BibitemShut
  {NoStop}%
\bibitem [{\citenamefont {\"Ohberg}\ and\ \citenamefont
  {Santos}(2002)}]{Santos:2002}%
  \BibitemOpen
  \bibfield  {author} {\bibinfo {author} {\bibfnamefont {P.}~\bibnamefont
  {\"Ohberg}}\ and\ \bibinfo {author} {\bibfnamefont {L.}~\bibnamefont
  {Santos}},\ }\href {\doibase 10.1103/PhysRevLett.89.240402} {\bibfield
  {journal} {\bibinfo  {journal} {Phys. Rev. Lett.}\ }\textbf {\bibinfo
  {volume} {89}},\ \bibinfo {pages} {240402} (\bibinfo {year}
  {2002})}\BibitemShut {NoStop}%
\bibitem [{\citenamefont {Pedri}\ \emph {et~al.}(2003)\citenamefont {Pedri},
  \citenamefont {Santos}, \citenamefont {\"Ohberg},\ and\ \citenamefont
  {Stringari}}]{Santos:2003}%
  \BibitemOpen
  \bibfield  {author} {\bibinfo {author} {\bibfnamefont {P.}~\bibnamefont
  {Pedri}}, \bibinfo {author} {\bibfnamefont {L.}~\bibnamefont {Santos}},
  \bibinfo {author} {\bibfnamefont {P.}~\bibnamefont {\"Ohberg}}, \ and\
  \bibinfo {author} {\bibfnamefont {S.}~\bibnamefont {Stringari}},\ }\href
  {\doibase 10.1103/PhysRevA.68.043601} {\bibfield  {journal} {\bibinfo
  {journal} {Phys. Rev. A}\ }\textbf {\bibinfo {volume} {68}},\ \bibinfo
  {pages} {043601} (\bibinfo {year} {2003})}\BibitemShut {NoStop}%
\bibitem [{\citenamefont {Peotta}\ and\ \citenamefont
  {Ventra}(2014)}]{Peotta:2014}%
  \BibitemOpen
  \bibfield  {author} {\bibinfo {author} {\bibfnamefont {S.}~\bibnamefont
  {Peotta}}\ and\ \bibinfo {author} {\bibfnamefont {M.~D.}\ \bibnamefont
  {Ventra}},\ }\href {\doibase 10.1103/PhysRevA.89.013621} {\bibfield
  {journal} {\bibinfo  {journal} {Phys. Rev. A}\ }\textbf {\bibinfo {volume}
  {89}},\ \bibinfo {pages} {013621} (\bibinfo {year} {2014})}\BibitemShut
  {NoStop}%
\bibitem [{\citenamefont {Choi}\ \emph {et~al.}(2015)\citenamefont {Choi},
  \citenamefont {Dunjko}, \citenamefont {Zhang},\ and\ \citenamefont
  {Olshanii}}]{Choi:2015}%
  \BibitemOpen
  \bibfield  {author} {\bibinfo {author} {\bibfnamefont {S.}~\bibnamefont
  {Choi}}, \bibinfo {author} {\bibfnamefont {V.}~\bibnamefont {Dunjko}},
  \bibinfo {author} {\bibfnamefont {Z.~D.}\ \bibnamefont {Zhang}}, \ and\
  \bibinfo {author} {\bibfnamefont {M.}~\bibnamefont {Olshanii}},\ }\href
  {\doibase 10.1103/PhysRevLett.115.115302} {\bibfield  {journal} {\bibinfo
  {journal} {Phys. Rev. Lett.}\ }\textbf {\bibinfo {volume} {115}},\ \bibinfo
  {pages} {115302} (\bibinfo {year} {2015})}\BibitemShut {NoStop}%
\bibitem [{\citenamefont {Campbell}\ \emph {et~al.}(2015)\citenamefont
  {Campbell}, \citenamefont {Gangardt},\ and\ \citenamefont
  {Kheruntsyan}}]{Campbell:2015}%
  \BibitemOpen
  \bibfield  {author} {\bibinfo {author} {\bibfnamefont {A.~S.}\ \bibnamefont
  {Campbell}}, \bibinfo {author} {\bibfnamefont {D.~M.}\ \bibnamefont
  {Gangardt}}, \ and\ \bibinfo {author} {\bibfnamefont {K.~V.}\ \bibnamefont
  {Kheruntsyan}},\ }\href {\doibase 10.1103/PhysRevLett.114.125302} {\bibfield
  {journal} {\bibinfo  {journal} {Phys. Rev. Lett.}\ }\textbf {\bibinfo
  {volume} {114}},\ \bibinfo {pages} {125302} (\bibinfo {year}
  {2015})}\BibitemShut {NoStop}%
\bibitem [{\citenamefont {Moritz}\ \emph {et~al.}(2003)\citenamefont {Moritz},
  \citenamefont {St\"oferle}, \citenamefont {K\"ohl},\ and\ \citenamefont
  {Esslinger}}]{Moritz:2003}%
  \BibitemOpen
  \bibfield  {author} {\bibinfo {author} {\bibfnamefont {H.}~\bibnamefont
  {Moritz}}, \bibinfo {author} {\bibfnamefont {T.}~\bibnamefont {St\"oferle}},
  \bibinfo {author} {\bibfnamefont {M.}~\bibnamefont {K\"ohl}}, \ and\ \bibinfo
  {author} {\bibfnamefont {T.}~\bibnamefont {Esslinger}},\ }\href {\doibase
  10.1103/PhysRevLett.91.250402} {\bibfield  {journal} {\bibinfo  {journal}
  {Phys. Rev. Lett.}\ }\textbf {\bibinfo {volume} {91}},\ \bibinfo {pages}
  {250402} (\bibinfo {year} {2003})}\BibitemShut {NoStop}%
\bibitem [{\citenamefont {Haller}\ \emph {et~al.}(2009)\citenamefont {Haller},
  \citenamefont {Gustavsson}, \citenamefont {Mark}, \citenamefont {Danzl},
  \citenamefont {R.~Hart},\ and\ \citenamefont {N\"{a}gerl}}]{Naagerl:2009}%
  \BibitemOpen
  \bibfield  {author} {\bibinfo {author} {\bibfnamefont {E.}~\bibnamefont
  {Haller}}, \bibinfo {author} {\bibfnamefont {M.}~\bibnamefont {Gustavsson}},
  \bibinfo {author} {\bibfnamefont {M.~J.}\ \bibnamefont {Mark}}, \bibinfo
  {author} {\bibfnamefont {J.~G.}\ \bibnamefont {Danzl}}, \bibinfo {author}
  {\bibfnamefont {G.~P.}\ \bibnamefont {R.~Hart}}, \ and\ \bibinfo {author}
  {\bibfnamefont {H.-C.}\ \bibnamefont {N\"{a}gerl}},\ }\href@noop {}
  {\bibfield  {journal} {\bibinfo  {journal} {Science}\ }\textbf {\bibinfo
  {volume} {325}},\ \bibinfo {pages} {1224} (\bibinfo {year}
  {2009})}\BibitemShut {NoStop}%
\bibitem [{\citenamefont {Fang}\ \emph {et~al.}(2014)\citenamefont {Fang},
  \citenamefont {Carleo}, \citenamefont {Johnson},\ and\ \citenamefont
  {Bouchoule}}]{Fang:2014}%
  \BibitemOpen
  \bibfield  {author} {\bibinfo {author} {\bibfnamefont {B.}~\bibnamefont
  {Fang}}, \bibinfo {author} {\bibfnamefont {G.}~\bibnamefont {Carleo}},
  \bibinfo {author} {\bibfnamefont {A.}~\bibnamefont {Johnson}}, \ and\
  \bibinfo {author} {\bibfnamefont {I.}~\bibnamefont {Bouchoule}},\ }\href
  {\doibase 10.1103/PhysRevLett.113.035301} {\bibfield  {journal} {\bibinfo
  {journal} {Phys. Rev. Lett.}\ }\textbf {\bibinfo {volume} {113}},\ \bibinfo
  {pages} {035301} (\bibinfo {year} {2014})}\BibitemShut {NoStop}%
\bibitem [{\citenamefont {Davis}\ \emph {et~al.}(2001)\citenamefont {Davis},
  \citenamefont {Morgan},\ and\ \citenamefont {Burnett}}]{Davis:2001b}%
  \BibitemOpen
  \bibfield  {author} {\bibinfo {author} {\bibfnamefont {M.~J.}\ \bibnamefont
  {Davis}}, \bibinfo {author} {\bibfnamefont {S.~A.}\ \bibnamefont {Morgan}}, \
  and\ \bibinfo {author} {\bibfnamefont {K.}~\bibnamefont {Burnett}},\ }\href
  {\doibase 10.1103/PhysRevLett.87.160402} {\bibfield  {journal} {\bibinfo
  {journal} {Phys. Rev. Lett.}\ }\textbf {\bibinfo {volume} {87}},\ \bibinfo
  {pages} {160402} (\bibinfo {year} {2001})}\BibitemShut {NoStop}%
\bibitem [{\citenamefont {Blakie}\ \emph {et~al.}(2008)\citenamefont {Blakie},
  \citenamefont {Bradley}, \citenamefont {Davis}, \citenamefont {Ballagh},\
  and\ \citenamefont {Gardiner}}]{Blakie:2008}%
  \BibitemOpen
  \bibfield  {author} {\bibinfo {author} {\bibfnamefont {P.~B.}\ \bibnamefont
  {Blakie}}, \bibinfo {author} {\bibfnamefont {A.~S.}\ \bibnamefont {Bradley}},
  \bibinfo {author} {\bibfnamefont {M.~J.}\ \bibnamefont {Davis}}, \bibinfo
  {author} {\bibfnamefont {R.~J.}\ \bibnamefont {Ballagh}}, \ and\ \bibinfo
  {author} {\bibfnamefont {C.~W.}\ \bibnamefont {Gardiner}},\ }\href {\doibase
  10.1080/00018730802564254} {\bibfield  {journal} {\bibinfo  {journal}
  {Advances in Physics}\ }\textbf {\bibinfo {volume} {57}},\ \bibinfo {pages}
  {363} (\bibinfo {year} {2008})}\BibitemShut {NoStop}%
\bibitem [{\citenamefont {Petrov}\ \emph {et~al.}(2000)\citenamefont {Petrov},
  \citenamefont {Shlyapnikov},\ and\ \citenamefont {Walraven}}]{Petrov:2000}%
  \BibitemOpen
  \bibfield  {author} {\bibinfo {author} {\bibfnamefont {D.~S.}\ \bibnamefont
  {Petrov}}, \bibinfo {author} {\bibfnamefont {G.~V.}\ \bibnamefont
  {Shlyapnikov}}, \ and\ \bibinfo {author} {\bibfnamefont {J.~T.~M.}\
  \bibnamefont {Walraven}},\ }\href {\doibase 10.1103/PhysRevLett.85.3745}
  {\bibfield  {journal} {\bibinfo  {journal} {Phys. Rev. Lett.}\ }\textbf
  {\bibinfo {volume} {85}},\ \bibinfo {pages} {3745} (\bibinfo {year}
  {2000})}\BibitemShut {NoStop}%
\bibitem [{\citenamefont {Richard}\ \emph {et~al.}(2003)\citenamefont
  {Richard}, \citenamefont {Gerbier}, \citenamefont {Thywissen}, \citenamefont
  {Hugbart}, \citenamefont {Bouyer},\ and\ \citenamefont
  {Aspect}}]{Richard:2003}%
  \BibitemOpen
  \bibfield  {author} {\bibinfo {author} {\bibfnamefont {S.}~\bibnamefont
  {Richard}}, \bibinfo {author} {\bibfnamefont {F.}~\bibnamefont {Gerbier}},
  \bibinfo {author} {\bibfnamefont {J.~H.}\ \bibnamefont {Thywissen}}, \bibinfo
  {author} {\bibfnamefont {M.}~\bibnamefont {Hugbart}}, \bibinfo {author}
  {\bibfnamefont {P.}~\bibnamefont {Bouyer}}, \ and\ \bibinfo {author}
  {\bibfnamefont {A.}~\bibnamefont {Aspect}},\ }\href {\doibase
  10.1103/PhysRevLett.91.010405} {\bibfield  {journal} {\bibinfo  {journal}
  {Phys. Rev. Lett.}\ }\textbf {\bibinfo {volume} {91}},\ \bibinfo {pages}
  {010405} (\bibinfo {year} {2003})}\BibitemShut {NoStop}%
\bibitem [{\citenamefont {van Amerongen}\ \emph {et~al.}(2008)\citenamefont
  {van Amerongen}, \citenamefont {van Es}, \citenamefont {Wicke}, \citenamefont
  {Kheruntsyan},\ and\ \citenamefont {van Druten}}]{Amerongen:2008}%
  \BibitemOpen
  \bibfield  {author} {\bibinfo {author} {\bibfnamefont {A.~H.}\ \bibnamefont
  {van Amerongen}}, \bibinfo {author} {\bibfnamefont {J.~J.~P.}\ \bibnamefont
  {van Es}}, \bibinfo {author} {\bibfnamefont {P.}~\bibnamefont {Wicke}},
  \bibinfo {author} {\bibfnamefont {K.~V.}\ \bibnamefont {Kheruntsyan}}, \ and\
  \bibinfo {author} {\bibfnamefont {N.~J.}\ \bibnamefont {van Druten}},\ }\href
  {\doibase 10.1103/PhysRevLett.100.090402} {\bibfield  {journal} {\bibinfo
  {journal} {Phys. Rev. Lett.}\ }\textbf {\bibinfo {volume} {100}},\ \bibinfo
  {pages} {090402} (\bibinfo {year} {2008})}\BibitemShut {NoStop}%
\bibitem [{\citenamefont {Hofferberth}\ \emph {et~al.}(2007)\citenamefont
  {Hofferberth}, \citenamefont {Lesanovsky}, \citenamefont {Fischer},
  \citenamefont {Schumm},\ and\ \citenamefont
  {Schmiedmayer}}]{Hofferberth:2007}%
  \BibitemOpen
  \bibfield  {author} {\bibinfo {author} {\bibfnamefont {S.}~\bibnamefont
  {Hofferberth}}, \bibinfo {author} {\bibfnamefont {I.}~\bibnamefont
  {Lesanovsky}}, \bibinfo {author} {\bibfnamefont {B.}~\bibnamefont {Fischer}},
  \bibinfo {author} {\bibfnamefont {T.}~\bibnamefont {Schumm}}, \ and\ \bibinfo
  {author} {\bibfnamefont {J.}~\bibnamefont {Schmiedmayer}},\ }\href
  {http://dx.doi.org/10.1038/nature06149} {\bibfield  {journal} {\bibinfo
  {journal} {Nature}\ }\textbf {\bibinfo {volume} {449}},\ \bibinfo {pages}
  {324} (\bibinfo {year} {2007})}\BibitemShut {NoStop}%
\bibitem [{\citenamefont {Jacqmin}\ \emph {et~al.}(2011)\citenamefont
  {Jacqmin}, \citenamefont {Armijo}, \citenamefont {Berrada}, \citenamefont
  {Kheruntsyan},\ and\ \citenamefont {Bouchoule}}]{Jacqmin:2011}%
  \BibitemOpen
  \bibfield  {author} {\bibinfo {author} {\bibfnamefont {T.}~\bibnamefont
  {Jacqmin}}, \bibinfo {author} {\bibfnamefont {J.}~\bibnamefont {Armijo}},
  \bibinfo {author} {\bibfnamefont {T.}~\bibnamefont {Berrada}}, \bibinfo
  {author} {\bibfnamefont {K.~V.}\ \bibnamefont {Kheruntsyan}}, \ and\ \bibinfo
  {author} {\bibfnamefont {I.}~\bibnamefont {Bouchoule}},\ }\href {\doibase
  10.1103/PhysRevLett.106.230405} {\bibfield  {journal} {\bibinfo  {journal}
  {Phys. Rev. Lett.}\ }\textbf {\bibinfo {volume} {106}},\ \bibinfo {pages}
  {230405} (\bibinfo {year} {2011})}\BibitemShut {NoStop}%
\bibitem [{\citenamefont {Armijo}\ \emph {et~al.}(2011)\citenamefont {Armijo},
  \citenamefont {Jacqmin}, \citenamefont {Kheruntsyan},\ and\ \citenamefont
  {Bouchoule}}]{Armijo:2011}%
  \BibitemOpen
  \bibfield  {author} {\bibinfo {author} {\bibfnamefont {J.}~\bibnamefont
  {Armijo}}, \bibinfo {author} {\bibfnamefont {T.}~\bibnamefont {Jacqmin}},
  \bibinfo {author} {\bibfnamefont {K.}~\bibnamefont {Kheruntsyan}}, \ and\
  \bibinfo {author} {\bibfnamefont {I.}~\bibnamefont {Bouchoule}},\ }\href
  {\doibase 10.1103/PhysRevA.83.021605} {\bibfield  {journal} {\bibinfo
  {journal} {Phys. Rev. A}\ }\textbf {\bibinfo {volume} {83}},\ \bibinfo
  {pages} {021605} (\bibinfo {year} {2011})}\BibitemShut {NoStop}%
\bibitem [{hea()}]{heat-transfer}%
  \BibitemOpen
  \href@noop {} {}\bibinfo {howpublished} {As is well known in acoustic
  physics, the contribution of heat transfer, estimated using a diffusion
  equation, to local changes of the energy per particle is negligible for
  long-wavelength deformations.}\BibitemShut {Stop}%
\bibitem [{\citenamefont {Kheruntsyan}\ \emph {et~al.}(2005)\citenamefont
  {Kheruntsyan}, \citenamefont {Gangardt}, \citenamefont {Drummond},\ and\
  \citenamefont {Shlyapnikov}}]{Kheruntsyan:2005}%
  \BibitemOpen
  \bibfield  {author} {\bibinfo {author} {\bibfnamefont {K.~V.}\ \bibnamefont
  {Kheruntsyan}}, \bibinfo {author} {\bibfnamefont {D.~M.}\ \bibnamefont
  {Gangardt}}, \bibinfo {author} {\bibfnamefont {P.~D.}\ \bibnamefont
  {Drummond}}, \ and\ \bibinfo {author} {\bibfnamefont {G.~V.}\ \bibnamefont
  {Shlyapnikov}},\ }\href {\doibase 10.1103/PhysRevA.71.053615} {\bibfield
  {journal} {\bibinfo  {journal} {Phys. Rev. A}\ }\textbf {\bibinfo {volume}
  {71}},\ \bibinfo {pages} {053615} (\bibinfo {year} {2005})}\BibitemShut
  {NoStop}%
\bibitem [{\citenamefont {Bouchoule}\ \emph {et~al.}(2007)\citenamefont
  {Bouchoule}, \citenamefont {Kheruntsyan},\ and\ \citenamefont
  {Shlyapnikov}}]{Bouchoule:2007}%
  \BibitemOpen
  \bibfield  {author} {\bibinfo {author} {\bibfnamefont {I.}~\bibnamefont
  {Bouchoule}}, \bibinfo {author} {\bibfnamefont {K.~V.}\ \bibnamefont
  {Kheruntsyan}}, \ and\ \bibinfo {author} {\bibfnamefont {G.~V.}\ \bibnamefont
  {Shlyapnikov}},\ }\href {\doibase 10.1103/PhysRevA.75.031606} {\bibfield
  {journal} {\bibinfo  {journal} {Phys. Rev. A}\ }\textbf {\bibinfo {volume}
  {75}},\ \bibinfo {pages} {031606} (\bibinfo {year} {2007})}\BibitemShut
  {NoStop}%
\bibitem [{foo()}]{footnote-arbitrary-omega}%
  \BibitemOpen
  \href@noop {} {}\bibinfo {howpublished} {We note that Eq. (4) and the scaling
  solutions (2)-(3) remain applicable for arbitrary time-variation of
  $\omega_1(t)$.}\BibitemShut {Stop}%
\bibitem [{sup()}]{sup}%
  \BibitemOpen
  \href@noop {} {}\bibinfo {howpublished} {See the Supplemental Material at
  http://link.aps.org/ supplemental/XXX, which contains additional proofs and
  clarifications of various aspects of the hydrodynamic solutions, as well as
  the details of $c$-field simulations.}\BibitemShut {Stop}%
\bibitem [{Ton()}]{Tonks-breathing-paper}%
  \BibitemOpen
  \href@noop {} {}\bibinfo {howpublished} {Y. Atas, I. Bouchoule, D. M.
  Gangardt, and K. V. Kheruntsyan, arXiv:1608.08720.}\BibitemShut {Stop}%
\bibitem [{sma()}]{small-amplitude}%
  \BibitemOpen
  \href@noop {} {}\bibinfo {howpublished} {For a small amplitude quench
  ($\epsilon\ll 1$), Eq. (\ref{eq-for-lambda-general}) with $\nu=1/2$ can be
  solved using the method of linearization [i.e., expanding $\lambda(t)$ as
  $\lambda(t)=1+\delta\lambda(t)$, with $\delta\lambda(t)\ll \lambda(t)$],
  yielding $\lambda(t)\simeq
  1+\frac{\epsilon}{3}-\frac{\epsilon}{3}\cos(\sqrt{3}\omega_1
  t)$.}\BibitemShut {Stop}%
\bibitem [{qBE()}]{qBEC-thermal-effects}%
  \BibitemOpen
  \href@noop {} {}\bibinfo {howpublished} {According to the quasicondensate
  equation of state, the last term in the HDE (\ref{eq:HDEb}),
  $\frac{1}{\rho}\partial_{x}P=g\partial_{x}\rho$, does not depend on
  temperature $T$ and is, in fact, the same as for a $T\!=\!0$ gas. Therefore,
  the HDEs for the density and velocity fields decouple from the equation for
  $s$ and consequently no finite-temperature effects are seen in the dynamics
  of $\rho(x,t)$ and $v(x,t)$.}\BibitemShut {Stop}%
\bibitem [{c-f()}]{c-field-c}%
  \BibitemOpen
  \href@noop {} {}\bibinfo {howpublished} {The regime of no doubling [bottom
  row in Fig. 1(a) and (b)], corresponding to a very small quench strength, is
  inaccessible in the $c$-field method due to large sampling
  errors.}\BibitemShut {Stop}%
\bibitem [{\citenamefont {Davis}\ \emph {et~al.}(2012)\citenamefont {Davis},
  \citenamefont {Blakie}, \citenamefont {van Amerongen}, \citenamefont {van
  Druten},\ and\ \citenamefont {Kheruntsyan}}]{Davis:2012}%
  \BibitemOpen
  \bibfield  {author} {\bibinfo {author} {\bibfnamefont {M.~J.}\ \bibnamefont
  {Davis}}, \bibinfo {author} {\bibfnamefont {P.~B.}\ \bibnamefont {Blakie}},
  \bibinfo {author} {\bibfnamefont {A.~H.}\ \bibnamefont {van Amerongen}},
  \bibinfo {author} {\bibfnamefont {N.~J.}\ \bibnamefont {van Druten}}, \ and\
  \bibinfo {author} {\bibfnamefont {K.~V.}\ \bibnamefont {Kheruntsyan}},\
  }\href {\doibase 10.1103/PhysRevA.85.031604} {\bibfield  {journal} {\bibinfo
  {journal} {Phys. Rev. A}\ }\textbf {\bibinfo {volume} {85}},\ \bibinfo
  {pages} {031604} (\bibinfo {year} {2012})}\BibitemShut {NoStop}%
\bibitem [{\citenamefont {Sinatra}\ \emph {et~al.}(2000)\citenamefont
  {Sinatra}, \citenamefont {Castin},\ and\ \citenamefont
  {Lobo}}]{sinatra2000monte}%
  \BibitemOpen
  \bibfield  {author} {\bibinfo {author} {\bibfnamefont {A.}~\bibnamefont
  {Sinatra}}, \bibinfo {author} {\bibfnamefont {Y.}~\bibnamefont {Castin}}, \
  and\ \bibinfo {author} {\bibfnamefont {C.}~\bibnamefont {Lobo}},\ }\href@noop
  {} {\bibfield  {journal} {\bibinfo  {journal} {Journal of Modern Optics}\
  }\textbf {\bibinfo {volume} {47}},\ \bibinfo {pages} {2629} (\bibinfo {year}
  {2000})}\BibitemShut {NoStop}%
\bibitem [{\citenamefont {Castin}\ \emph {et~al.}(2000)\citenamefont {Castin},
  \citenamefont {Dum}, \citenamefont {Mandonnet}, \citenamefont {Minguzzi},\
  and\ \citenamefont {Carusotto}}]{Castin:2000}%
  \BibitemOpen
  \bibfield  {author} {\bibinfo {author} {\bibfnamefont {Y.}~\bibnamefont
  {Castin}}, \bibinfo {author} {\bibfnamefont {R.}~\bibnamefont {Dum}},
  \bibinfo {author} {\bibfnamefont {E.}~\bibnamefont {Mandonnet}}, \bibinfo
  {author} {\bibfnamefont {A.}~\bibnamefont {Minguzzi}}, \ and\ \bibinfo
  {author} {\bibfnamefont {I.}~\bibnamefont {Carusotto}},\ }\href {\doibase
  10.1080/09500340008232189} {\bibfield  {journal} {\bibinfo  {journal}
  {Journal of Modern Optics}\ }\textbf {\bibinfo {volume} {47}},\ \bibinfo
  {pages} {2671} (\bibinfo {year} {2000})}\BibitemShut {NoStop}%
\bibitem [{\citenamefont {Dennis}\ \emph {et~al.}(2013)\citenamefont {Dennis},
  \citenamefont {Hope},\ and\ \citenamefont {Johnsson}}]{Dennis:2012}%
  \BibitemOpen
  \bibfield  {author} {\bibinfo {author} {\bibfnamefont {G.~R.}\ \bibnamefont
  {Dennis}}, \bibinfo {author} {\bibfnamefont {J.~J.}\ \bibnamefont {Hope}}, \
  and\ \bibinfo {author} {\bibfnamefont {M.~T.}\ \bibnamefont {Johnsson}},\
  }\href@noop {} {\bibfield  {journal} {\bibinfo  {journal} {Computer Physics
  Communications}\ }\textbf {\bibinfo {volume} {184}},\ \bibinfo {pages} {201}
  (\bibinfo {year} {2013})}\BibitemShut {NoStop}%
\end{thebibliography}

%

\onecolumngrid\newpage



\section*{Supplementary material}

I. Bouchoule,[1] S. S. Szigeti,[2,3] M. J. Davis,[2] and K. V. Kheruntsyan[2]

[1]*Laboratoire Charles Fabry, Institut d'Optique, CNRS, Univesité Paris Sud 11, 2 Avenue Augustin Fresnel, F-91127 Palaiseau Cedex, France*
[2]*University of Queensland, School of Mathematics and Physics, Brisbane, Queensland 4072, Australia*
[3]*ARC Centre of Excellence for Engineered Quantum Systems, University of Queensland, Brisbane, Queensland 4072, Australia*

In this supplemental material we provide further details on the hydrodynamic scaling solutions, as well as a brief description and further results of our *c*-field simulations.

## I. SCALING SOLUTIONS IN THE HYDRODYNAMIC APPROACH

### A. Ideal gas regime.

For a uniform ideal gas (either bosonic or fermionic) at temperature $T$ and 1D density $\rho = N/L$, where $N$ is the number of particles and $L$ is length of the confinement box, the only two length scales are the mean interparticle separation $\rho^{-1}$ and the thermal de Broglie wavelength $\Lambda_T = \sqrt{2\pi\hbar^2/(mk_BT)}$; the corresponding energy scales are $\hbar^2\rho^2/m$ and $k_BT$. Using the thermodynamic definition of the 1D pressure, $P = (\partial U/\partial L)_s$, where $U$ is the internal energy, one can apply simple dimensional analysis to write down the equations of state for $P$ and $s$:

$$P = k_BT\rho \; F(\hbar^2\rho^2/mk_BT), \quad (S1)$$
$$s/k_B = G(mk_BT/\hbar^2\rho^2). \quad (S2)$$

Here $F$ and $G$ are dimensionless functions of the only dimensionless parameter—the ratio of the two energy scales. With this choice of expression for $P$, the classical ideal gas law in the high temperature limit is recovered with $F \simeq 1$. For a highly degenerate ideal Fermi gas, on the other hand, the equation of state $P = \hbar^2\pi^2\rho^3/(3m)$ is recovered with $F \simeq (\pi^2/3)(\hbar^2\rho^2/mk_BT)$.

Now consider a confinement quench of the gas. Applying the general functional forms of $P$ and $s$ to small (locally uniform) slices of the gas, it can be shown by direct substitution that the scaling solutions, Eqs. (2) and (3) of the main text, satisfy Eqs. (1a) and (1c). For Eq. (1b), first note that Eq. (S1) together with the scaling solutions imply that $P(x,t) = P_0(x/\lambda)/\lambda^3$, where $P_0(x) \equiv P(x,0)$. Since Eq. (1b) is assumed true at this initial time, then $\partial_x P_0 = -\rho_0(x)\partial_x V(x,0) = -m\omega_0^2 x\rho_0(x)$. Together with the scaling solutions and Eq. (4) of the main text, this relation is sufficient to show that Eq. (1b) is true for all times.

### B. Equilibrium momentum distribution of a uniform 1D quasicondensate.

Here we outline the derivation of the Lorentzian shape of the equilibrium momentum distribution of a uniform 1D quasicondensate (for a more detailed derivation, we refer the reader to Refs. [1, 2]). For a uniform and hence translationally invariant system, the momentum distribution is given by the Fourier transform of the first-order correlation function, $G^{(1)}(x) = \langle \hat{\Psi}^\dagger(x)\hat{\Psi}(0)\rangle$, where $\hat{\Psi}(x)$ is the bosonic field operator. In the quasicondensate regime, corresponding to $T \ll \sqrt{\gamma} \; \hbar^2\rho^2/(2mk_B)$ [3] (with $\rho$ being the uniform density and $\gamma \equiv mg/\hbar^2\rho$), the first-order correlation function is dominated by the long-wavelength (low-energy) excitations whose Hamiltonian reduces (using the density-phase representation of the field operator, $\hat{\Psi}(x) = \sqrt{\rho + \delta\hat{\rho}(x)}e^{i\hat{\phi}(x)}$) to the Luttinger liquid form [2, 4]

$$\hat{H}_L = \int dx \left[\frac{\hbar^2\rho}{2m}(\partial_x\hat{\phi})^2 + \frac{g}{2}(\delta\hat{\rho})^2\right]. \quad (S3)$$

Here, $\delta\hat{\rho}(x)$ is the operator describing the density fluctuations, canonically conjugate to the phase operator $\hat{\phi}(x)$.

In evaluating $G^{(1)}(x)$ we note that the density fluctuations are small in the quasicondensate regime and can be completely neglected as long as the relative distances of interest are much larger than the healing length $\xi = \hbar/\sqrt{mg\rho}$ [1, 2, 5, 6]. As the Luttinger liquid Hamiltonian (S3) is quadratic in $\hat{\phi}$, the correlation function $G^{(1)}(x) = \rho\langle e^{i(\hat{\phi}(x)-\hat{\phi}(0))}\rangle$ can be expressed through the mean-square fluctuations of the phase via Wick's theorem:

$$G^{(1)}(x) = \rho e^{-\frac{1}{2}\langle[\hat{\phi}(x)-\hat{\phi}(0)]^2\rangle}. \quad (S4)$$

Denoting the Fourier component of $\hat{\phi}(x)$ at wavevector $k$ via $\hat{\phi}_k$, the corresponding term in the expectation value of the Hamiltonian is given by $\int \frac{dk}{2\pi} L\rho\hbar^2k^2\langle|\hat{\phi}_k|^2\rangle/(2m)$, where $L$ is the length of the uniform system. The relevant modes contributing to $G^{(1)}(x)$ are highly populated, such that a classical field picture is sufficient. The energy per quadratic degree of freedom is thus given by $k_BT/2$ and therefore $\langle|\hat{\phi}_k|^2\rangle = mk_BT/(L\rho\hbar^2k^2)$ [7]. Using this result one can then show, after a little algebra, that Eq. (S4) yields

$$G^{(1)}(x) \simeq \rho e^{-|x|/2l_\phi}, \quad (|x| \gg \xi), \quad (S5)$$

where $l_\phi \equiv \hbar^2\rho/(mk_BT)$. The Fourier transform of this exponentially decaying correlation function gives the Lorentzian momentum distribution, $\bar{n}(k) = (2\rho l_\phi/\pi)/\left[1+(2l_\phi k)^2\right]$, used in the main text.

### C. Scaling solution for the temperature in the quasicondensate regime.

In order to calculate $n(k,t)$ using the hydrodynamic scaling solutions, Eqs. (2) and (4) of the main text, we first need to determine how the temperature of the gas and hence the phase correlation length $l_\phi = \hbar^2\rho/(mk_BT)$ evolves during the breathing oscillations. To do this, we first note that the energy of the $j$th phonon mode in a quantization box of length $L$ is given by $E_j = \hbar k_j c$, where $k_j = \frac{2\pi}{L}j$ is the phonon wave vector and $c = \sqrt{(\partial P/\partial \rho)/m}$ is the speed of sound. Using $P \propto \rho^2$, we find that $c$ scales as $c \propto \rho^{1/2}$ and therefore $E_j \propto L^{-1}\rho^{1/2} \propto \rho^{3/2}$. Consider now an adiabatic compression/decompression cycle for a uniform slice of the gas confined to a box of length $L$. Such a compression does not change the mean occupation number of the mode $j$. The mode occupation is given by $n_j \simeq k_BT/E_j$ in the long-wavelength limit and scales as $n_j \propto T/\rho^{3/2}$, whereas $\rho \propto \lambda^{-1}$ according to the scaling solution, Eq. (2) of the main text. Therefore, during the adiabatic breathing oscillations, the temperature $T(t)$ evolves from the initial value $T(0) \equiv T_0$ to $T(t) = T_0/\lambda^{3/2}$, i.e., Eq. (3) of the main text with $\nu = 1/2$.

### D. Evolution of the momentum distribution for an ideal gas.

Applying the hydrodynamic approach and Eq. (7) of the main text to the ideal gas regime, we first note that the momentum distribution of a uniform ideal gas (normalized to $\int dk\, \bar{n}(k) = \rho$) is given by $\bar{n}(\rho,s;k) = \mathcal{N}((\hbar^2k^2/2m - \mu)/k_BT)$, where $\mathcal{N}$ is a dimensionless function whose expression depends on the quantum statistics [8]. Since $\mu/k_BT$ is a function of $s$ (in the sense of a thermodynamic equation of state), which itself is a function of $mk_BT/\hbar^2\rho^2$ [see Eq. (S2)], one can assert that $\mu/(k_BT) = \mathcal{G}(mk_BT/\hbar^2\rho^2)$, where $\mathcal{G}$ is a dimensionless function. Then, the scaling solutions (2) and (3) of the main text imply that $\mu(x,t)/k_BT(t) = \mathcal{G}[mk_BT(t)/\hbar^2\rho^2(x,t)] = \mathcal{G}[mk_BT_0/\hbar\rho_0^2(x/\lambda)] \equiv \mathcal{G}_0(x/\lambda)$, or

$$\mu(x,t) = \mu(x/\lambda,0)/\lambda^2 = [\mu_0 - \tfrac{1}{2}m\omega_0^2(x/\lambda)^2]/\lambda^2, \quad \text{(S6)}$$

where $\mu_0$ is the initial chemical potential in the trap center. Substituting $\bar{n}(\rho,s;k)$ along with this expression for $\mu(x,t)$ into Eq. (6) of the main text, and changing variables to $\tilde{x} = \alpha x - \hbar k\lambda\dot{\lambda}/(m\omega_0^2\alpha)$, gives

$$n(k,t) = n_0(k/\alpha)/\alpha, \quad \text{(S7)}$$

where $n_0(k)$ is the initial momentum distribution of the trapped gas and $\alpha^2 = (\omega_0^2 + \lambda^2\dot{\lambda}^2)/(\lambda\omega_0)^2$. Using Eq. (5) of the main text, we can explicitly write $\alpha$ as

$$\alpha = \sqrt{[1+\epsilon\cos^2(\omega_1 t)]/(1+\epsilon)}, \quad \text{(S8)}$$

which implies that the momentum distribution of a finite-temperature ideal gas in the hydrodynamic limit oscillates at $\omega_B = 2\omega_1$ and *never* displays frequency doubling. One thus recovers the expected behaviour for an ideal gas, due to the position-momentum symmetry of the underlying harmonic oscillator Hamiltonian. The fact that this result is reproduced within the hydrodynamic approach is a result of a "fortuitous" exact cancelation of the effect of the hydrodynamic velocity field by the thermal component.

### E. Scaling of the frequency doubling crossover $A_{\mathrm{cr}}$ with the quench strength $\epsilon$.

In this section we make qualitative arguments that provide an understanding of the dominant oscillation regimes in the dynamics of the momentum distribution of a quasi-condensate, and derive an approximate scaling of $A_{\mathrm{cr}}$ with $\epsilon$. Let us first introduce typical momentum scales involved in the dynamics. For small-amplitude oscillations, corresponding to $\epsilon \ll 1$, the scaling parameter $\lambda$ oscillates as $\lambda(t) \simeq 1 + \frac{\epsilon}{3} - \frac{\epsilon}{3}\cos(\sqrt{3}\omega_1 t)$ and therefore the magnitude of $\dot{\lambda}$ is of the order of $\sim \epsilon/\sqrt{3}$. This means that the characteristic hydrodynamic momentum, which can be estimated as $\bar{k}_{\mathrm{h}} \sim mX_0\dot{\lambda}/\hbar$ from Eq. (7) of the main text, and which can be rewritten as $\bar{k}_{\mathrm{h}} \sim mX_0\epsilon\omega_1/\hbar \sim (\omega_1/\omega_0)A\epsilon/l_\phi^{(0)}$, is of the order of $\bar{k}_{\mathrm{h}} \sim A\epsilon/l_\phi^{(0)}$ for $\epsilon \ll 1$. Compared to this, the characteristic thermal momentum during the compression/decompression cycle oscillates above $\bar{k}_{\mathrm{th}} \sim 1/l_\phi^{(0)}$ with an amplitude variation of $\delta\bar{k}_{\mathrm{th}} \sim \epsilon/l_\phi^{(0)} \ll \bar{k}_{\mathrm{th}}$.

For $A\epsilon \gg 1$, the characteristic hydrodynamic momentum is much larger than both the characteristic thermal momentum and its variation, $\bar{k}_{\mathrm{h}} \gg \bar{k}_{\mathrm{th}}, \delta\bar{k}_{\mathrm{th}}$. Then, as long as one is interested in momenta of the order of $\bar{k}_{\mathrm{h}}$, the function $\bar{n}$ in Eq. (6) of the main text can be approximated by a $\delta$-function, and therefore the breathing oscillations of the momentum distribution will be dominated by the hydrodynamic phenomenon of frequency doubling.

In the opposite regime of $A\epsilon \ll 1$, the characteristic hydrodynamic momentum is much smaller than the characteristic thermal momentum, $\bar{k}_{\mathrm{h}} \ll \bar{k}_{\mathrm{th}}$, and therefore the above approximation, which neglects the effect of the width of $\bar{n}$ in Eq. (6) of the main text, breaks down. In this case, the contribution of the hydrodynamic momenta to $n(k,t)$ can instead be estimated via a Taylor series of $\bar{n}$ as powers of $k_{\mathrm{h}}(x,t) = mv(x,t)/\hbar$. In this series, the contribution of the first-order term to the integral vanishes because $\partial\bar{n}/\partial k$ is an even function of $x$, whereas $k_{\mathrm{h}}(x,t)$ is odd; therefore, the hydrodynamic velocity field has no effect on $n(k,t)$ in this order. The leading-order correction in $n(k,t)$ thus comes from the second-order derivative term, proportional to $k_{\mathrm{h}}^2$. To estimate this correction, let us consider the typical variations of the peak value of the momentum distribution $n(0,t)$, which we denote via $\delta n$, induced solely by the hydrodynamic momenta. Using $\partial^2\bar{n}/\partial k^2|_{k=0} \sim \rho_0(0)(l_\phi^{(0)})^3$ in Eq. (6), one can find that $\delta n \sim -X_0\rho_0(0)l_\phi^{(0)3}\bar{k}_{\mathrm{h}}^2$, or

$\delta n \sim -X_0 \rho_0(0) l_\phi^{(0)} (A\epsilon)^2$, where we have used $\bar{k}_{\rm h} \sim A\epsilon/l_\phi^{(0)}$. Since the peak value $n(0,t)$ is inversely proportional to the characteristic momentum width $W$, the hydrodynamic contribution to the typical change ($\delta W$) of the momentum width fulfils $\delta W/W \simeq -\delta n/n(0,0)$. Using $W \sim \bar{k}_{\rm th} \sim 1/l_\phi^{(0)}$ and $n(0,0) \sim \rho_0(0) l_\phi^{(0)} X_0$, we then find $\delta W \sim (A\epsilon)^2/l_\phi^{(0)}$. Comparing now this result with the typical variation of the thermal momentum width $\delta \bar{k}_{\rm th} \sim \epsilon/l_\phi^{(0)}$, we can conclude that the hydrodynamic contribution $\delta W \sim (A\epsilon)^2/l_\phi^{(0)}$ will dominate the thermal contribution if $A \gg 1/\sqrt{\epsilon}$; in this case, one would still observe the phenomenon of frequency doubling. If, on the other hand, $A \ll 1/\sqrt{\epsilon}$, the breathing oscillations will be dominated by the variations of thermal momenta and no frequency doubling will be observed. Accordingly, one can expect the crossover from doubling to no doubling to occur at $A_{\rm cr} \propto 1/\sqrt{\epsilon}$, with the proportionality factor to be found from the numerical results. From Fig. 1(c) of the main text we see that, in the relevant region of $A \gtrsim 5$, $A_{\rm cr}$ scales as $1/\sqrt{\epsilon}$ as expected. Since $A \gg 1$ in the quasicondensate regime, the frequency doubling crossover requires very small quench strengths $\epsilon$.

## II. NUMERICAL SIMULATIONS USING THE $C$-FIELD METHODOLOGY

### A. The $c$-field method

The $c$-field (or classical field) method is a proven approach to studying the equilibrium properties and dynamics of degenerate Bose gases at finite temperature [9]. The crux of the technique is to treat the quantum Bose field $\hat{\psi}(x,t)$ as a classical field $\psi_{\rm C}(x,t)$, thus ignoring the discrete nature of the particles that make up the field. The classical field approximation captures many features of weakly-interacting 1D Bose gases. For instance, for thermal equilibrium configurations it correctly describes the crossover from the ideal Bose gas regime to the quasicondensate regime [10].

The energy functional of a classical Bose field confined in a harmonic potential is

$$E(\{\psi_{\rm C}\}) = \int dx\, \mathcal{E}(x), \quad (S9)$$

where

$$\mathcal{E}(x) = \frac{\hbar^2}{2m}\left|\frac{\partial \psi_{\rm C}(x)}{\partial x}\right|^2 + \frac{1}{2}m\omega_0^2 x^2 |\psi_{\rm C}(x)|^2 + \frac{g}{2}|\psi_{\rm C}(x)|^4 - \mu|\psi_{\rm C}(x)|^2. \quad (S10)$$

Configurations corresponding to thermal equilibrium are obtained from the Gibbs ensemble: the probability of a field configuration $\psi_{\rm C}(x)$ is proportional to $\exp(-E(\{\psi_{\rm C}\})/k_B T)$. Here $\mu$ is the chemical potential that fixes the mean particle number.

A convenient method to sample configurations from the Gibbs ensemble is to integrate the projected stochastic Gross-Pitaevskii equation (SPGPE) for times long enough that the memory of the initial state is lost. The SPGPE is

$$d\psi_{\rm C}(x,t) = \mathcal{P}_{\rm C}\bigg\{-\frac{i}{\hbar}\mathcal{L}_{\rm C} + \kappa_{\rm th}(\mu - \mathcal{L}_{\rm C})\psi_{\rm C}(x,t)dt + \sqrt{2\kappa_{\rm th} T}dW(x,t)\bigg\}, \quad (S11)$$

where

$$\mathcal{L}_{\rm C}\psi_{\rm C} = \left[-\frac{\hbar}{2m}\frac{\partial^2}{\partial x^2} + \frac{1}{2}m\omega_0^2 x^2 + g|\psi_{\rm C}|^2\right]\psi_{\rm C}, \quad (S12)$$

$dW(x,t)$ is uncorrelated complex white noise satisfying $\langle dW^*(x,t)dW(x,t')\rangle = \delta(x-x')dt$, and $\mathcal{P}_{\rm C}\{\cdot\}$ is the projector onto the computational basis. The value of the rate $\kappa_{\rm th}$ has no consequence for the equilibrium configurations, and hence can be chosen for numerical convenience [11].

At this point we make some comments about the classical field model to present its limitations and provide some physical insight. For high energy modes of the classical field, the interaction energy can be neglected and the energy functional can be approximated by that for noninteracting particles. Given a mode of energy $\epsilon_m$, the classical field model predicts an energy of $k_B T$, and thus a mean occupation number of $n_m = k_B T/(\epsilon_m - \mu)$. Although this expression is a good approximation to the Bose-Einstein distribution for $n_m \gg 1$, it overestimates the population for $n_m \lesssim 1$, where the Maxwell-Boltzmann distribution is a more appropriate model. In 2D and 3D, this overestimation leads to an ultraviolet divergence of the field density $|\psi_{\rm C}|^2$ at a fixed temperature. To overcome this problem, one can introduce an energy cutoff: higher-energy modes are treated as an ideal Bose gas [12–15], while the classical field is restricted to the low-energy modes. With this separation of the field into a classical and a quantum part (in the sense that the discreteness of particles is important), Eq. (S11) has a physical meaning: it represents the "real" time evolution of the field with $\kappa_{\rm th}$ an effective collision rate that quantifies the thermal and diffusive damping experienced by particles in the classical field region due to the thermal reservoir (i.e. the quantum region) [9].

Although the inclusion of an energy cutoff is crucial in higher dimensions in order to prevent a divergence of the atomic density, its role is less crucial in 1D. In particular, the 1D classical field predictions for the atomic density do not diverge, even in absence of an energy cutoff, and are quantitatively correct for degenerate gases [16]. The results we present are correctly described solely by classical field theory without a cutoff: for a given peak linear density and temperature, the results do not depend on the cutoff once it is large enough. The cutoff merely determines the size of the basis used for the numerical calculations.

Our simulations were performed within the Hermite-Gauss basis, which is the single-particle eigenbasis of an harmonically-confined ideal gas, and therefore represents the natural, most computationally efficient basis (see Eq. (39) of [17] for an explicit expression of the SPGPE in the Hermite-Gauss basis). Since these eigenstates well-approximate the higher-energy, sparsely populated modes, simulation in the

Hermite-Gauss basis imposes an energy cutoff that is in direct proportion to the basis size. As discussed above, the variation of this energy cutoff makes little difference to the resulting equilibrium states.

After generating the equilibrium ensemble, the simulations proceed by quenching the trapping frequency $\omega_0 \to \omega_1$ and evolving the ensemble in time using the simpler projected Gross-Pitaevskii equation (PGPE) [which can be obtained by setting $\kappa_{th} = 0$ in Eq. (S11)] [18, 19]. This equation then conserves energy and number of particles. As pointed out above, this classical field approximation fails to correctly capture the behavior of high energy modes. In contrast to the equilibrium case, this can affect the dynamics, and in particular the damping rates, and the details can be sensitive to the cutoff [20, 21]. Nevertheless, nonequilibrium properties have been studied with some success using this type of classical field approximation, including the collective oscillations of Bose gases [21, 22]. In our case, the results of dynamics that we present here are found not to depend strongly on the choice of the energy cutoff, demonstrating that the role of the higher-energy states is relatively unimportant.

For consistency, we chose the same energy cutoff for both the PGPE evolution after the confinement quench and the SPGPE that generates the initial condition. However, the Hermite-Gauss modes that represent the single-particle eigenbasis depend upon the trapping frequency. This implies that the number of Hermite-Gauss modes used for the PGPE evolution, $\tilde{M}_{\text{cut}}$, is related to the number of modes in the SPGPE evolution $M_{\text{cut}}$ via

$$\tilde{M}_{\text{cut}} = \left\lfloor \frac{\omega_1}{\omega_0}(M_{\text{cut}} + \tfrac{1}{2}) - \tfrac{1}{2} \right\rfloor, \quad (S13)$$

where $\lfloor x \rfloor$ denotes the integer component of $x$.

Figure 1 illustrates PGPE evolution of the position density and momentum distributions after a confinement quench in three different parameter regimes. Although the density undergoes breathing oscillations at frequency $\omega_B \simeq \sqrt{3}\omega_1$ in all three cases, the momentum distribution exhibits frequency doubling (top row), a crossover between quasi-doubling and no doubling (middle row), and no frequency doubling (bottom row). This is consistent with the breathing oscillations predicted by our finite-temperature hydrodynamic theory within the quasicondensate regime (see main text).

### B. Details for comparison with finite-temperature hydrodynamic theory in the quasicondensate regime

The thermal equilibrium properties of a harmonically trapped Bose gas at temperature $T_0$ and peak density $\rho_0(0)$ can be parametrized by the three dimensionless quantities $\gamma_0 = mg/[\hbar^2 \rho_0(0)]$, $t_0 = 2\hbar^2 k_B T_0/(mg^2)$, and $\tilde{\omega} \equiv (l_{\text{HO}}/\xi_0) = \hbar\omega_0/[g\rho_0(0)]^2$, where $l_{\text{HO}} = \sqrt{\hbar/(m\omega_0)}$ and $\xi_0 = \hbar/\sqrt{mg\rho_0(0)}$ are the harmonic oscillator length scale and healing length, respectively. However, within the classical field approximation only two parameters are required since features on the order of mean interparticle separation $1/\rho_0(0)$ are neglected. Specifically, if the classical field is scaled by $\psi_0 = [mk_B^2 T_0^2/(\hbar^2 g)]^{1/6}$, the length scale by $x_0 = (\hbar^4/(m^2 g k_B T_0))^{1/3}$, and the trapping frequency by $\bar{\omega} = (k_B T_0 \sqrt{mg}/\hbar)^{2/3}/\hbar$, then the thermal action $E(\{\psi_\mathbf{C}\})/k_B T$ [see Eqs. (S9) and (S10)], which determines the grand canonical partition function, depends only on the dimensionless parameters $\eta_0 = [\hbar^2/(mg^2 k_B^2 T_0^2)]^{1/3}\mu$ and $\omega_0/\bar{\omega}$. Equivalently, the gas can be described by the dimensionless parameters $\chi_0 = k_B T/[\hbar\rho_0(0)\sqrt{g\rho_0(0)/m}]$ and $\tilde{\omega} = \hbar\omega/[g\rho_0(0)]$; field correlation functions of order $q$ have previously been shown to depend only on $\chi_0$ and $\tilde{\omega}$, provided they are scaled to $\rho_0(0)^q$ and the lengths are scaled to $\hbar^2\rho_0/(mk_B T)$ (cf. [10, 23] which investigated the parameter dependence within the classical field approximation for a *uniform* Bose gas). For sufficiently weak trapping frequencies, $\tilde{\omega} \to 0$ and drops out of the problem. This occurs if the size of the atomic cloud is much larger than other microscopic correlation lengths of the gas, therefore implying that the local density approximation (LDA) is valid.

Consider now the post-quench dynamics investigated in this paper. They are *a priori* parametrized by the dimensionless parameters $\gamma_0$, $A = \sqrt{8}/\chi_0 = \sqrt{8}/(\gamma_0^{3/2} t_0)$, $\epsilon$, and $\tilde{\omega}$. Within the classical field approximation, if we rescale the PGPE [cf. Eqs. (S11) and (S12)] as done for the equilibrium case, we find that $\gamma_0$ drops out. Additionally, we chose parameters for our $c$-field simulations such that the dynamics only depend upon $A$ and $\epsilon$. That is, we required a sufficiently weak $\tilde{\omega}$ such that the size of the cloud was always much larger than the typical correlation length $l_\phi^{(0)} = \hbar^2 \rho_0(0)/(mk_B T_0)$ (recall that $l_\phi^{(0)}$ is the typical phase correlation length of the gas, itself larger or on the order of the density-density correlation length). Similarly to the equilibrium case, one then expects a dynamical LDA to be valid, ensuring that the parameter $\tilde{\omega}$ is irrelevant. This further fulfilled the high temperature condition required for the (S)PGPE, whilst still ensuring that the number of modes was numerically tractable.

In order to compute $A_{\text{cr}}$ for a given quench strength $\epsilon$, we fixed $t_0$ and varied $A$. For each $A$, $K$ was extracted by fitting $B_1 \exp(-b_1 t)[\sqrt{K}\cos(\nu t) - \sqrt{1-K}\cos(2\nu t)] + B_2 \exp(-b_2 t)$ to the HWHM of the momentum distribution at each time point $t$ (here $B_1$, $B_1$, $b_1$, $b_2$, $\nu$, and $K$ are free fitting parameters); see the red curves of Fig. 1 for example fits. The rates $b_1$ and $b_2$ account for the damping present in the PGPE evolution, which is absent from our hydrodynamic theory. This gave a dataset $(A, K)$; the point $A_{\text{cr}}$ where $K = 1/2$ was determined by fitting $\frac{1}{2}[\tanh[a(A^{-2/3} - A_{\text{cr}}^{-2/3})] + 1]$ to this dataset, with free fitting parameters $a$ and $A_{\text{cr}}$.

The equilibrium momentum distributions generated by the SPGPE differ slightly to those generated from Eq. (10) of the main text. This is not unexpected; various assumptions that go into Eq. (10) (such as a Thomas-Fermi density profile and the LDA) are relaxed in the SPGPE. A fairer comparison to our finite-temperature hydrodynamic theory is therefore obtained by fitting the equilibrium momentum distribution predicted by Eq. (10) of the main text to the $c$-field equilibrium momentum distribution (with $A$ and $\hbar\omega_0/[g\rho_0(0)]$ as free parameters). This shifts the dataset $(A, K) \to (A', K)$; the $c$-field

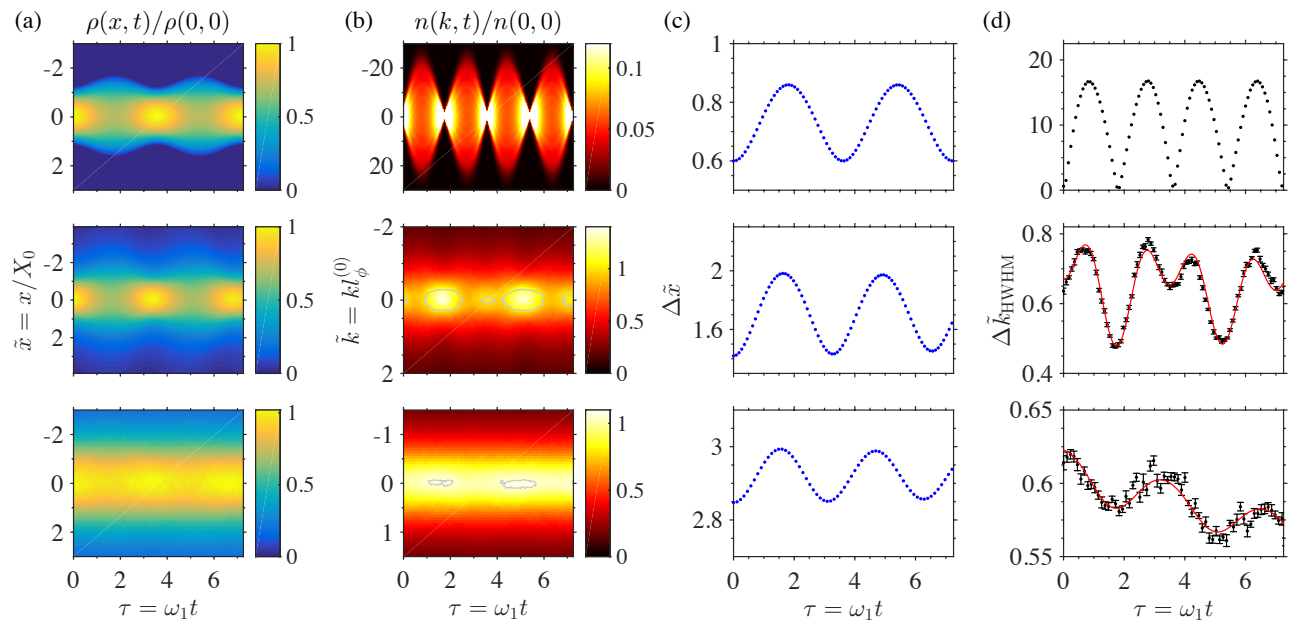

FIG. 1. (Color online). Results of PGPE simulations illustrating the breathing mode oscillations of a harmonically trapped, weakly-interacting 1D Bose gas after a confinement quench. (a) Density; (b) momentum distribution; (c) rms width $\Delta \tilde{x}$ of the density; and (d) half-width-at-half-maximum $\Delta \tilde{k}_{\mathrm{HWHM}}$ of the momentum distribution; all as a function of the dimensionless time $\tau = \omega_1 t$. The three examples correspond, respectively, to: $\epsilon = 0.563$, $A = 104$, $t_0 = 10^6$ and $\hbar\omega_0/[g\rho_0(0)] = 3.0 \times 10^{-3}$ – top row; $\epsilon = 0.778$, $A \approx A_{\mathrm{cr}} = 2.3$, $t_0 = 10^3$, and $\hbar\omega_0/[g\rho_0(0)] = 0.07$ – middle row; and $\epsilon = 0.1$, $A = 1.15$, $t_0 = 10^6$, and $\hbar\omega_0/[g\rho_0(0)] = 0.1$ – bottom row. In (d) the black points are c-field data, with the error bars indicating 95% confidence intervals, whilst the solid red curves are fits of the form $B_1 \exp(-b_1 t)[\sqrt{K}\cos(\nu t) - \sqrt{1-K}\cos(2\nu t)] + B_2 \exp(-b_2 t)$.

values of $A_{\mathrm{cr}}$ reported in Fig. 1 of the main text are computed with respect to this latter dataset.

[1] C. Mora and Y. Castin, Phys. Rev. A **67**, 053615 (2003).
[2] M. A. Cazalilla, Journal of Physics B: Atomic, Molecular and Optical Physics **37**, S1 (2004).
[3] K. V. Kheruntsyan, D. M. Gangardt, P. D. Drummond, and G. V. Shlyapnikov, Phys. Rev. A **71**, 053615 (2005).
[4] F. D. M. Haldane, Phys. Rev. Lett. **47**, 1840 (1981).
[5] A. G. Sykes, D. M. Gangardt, M. J. Davis, K. Viering, M. G. Raizen, and K. V. Kheruntsyan, Phys. Rev. Lett. **100**, 160406 (2008).
[6] P. Deuar, A. G. Sykes, D. M. Gangardt, M. J. Davis, P. D. Drummond, and K. V. Kheruntsyan, Phys. Rev. A **79**, 043619 (2009).
[7] Since $\hat{\phi}(x)$ is a real function in the classical field picture, one should, in fact, expand on real sinusoidal functions; see: I. Bouchoule, N. J. van Druten, and C. I. Westbrook, in *Atom Chips*, Eds. J. Reichel and V. Vuletic (Wiley-VCH, Weinheim, Germany, 2011).
[8] Explicitly, the dimensionless function $\mathcal{N}$ for a 1D Bose or Fermi gas is given by $\mathcal{N}((\hbar^2 k^2/2m - \mu)/k_B T) = \frac{1}{2\pi}[e^{(\hbar^2 k^2/2m - \mu)/k_B T} \mp 1]^{-1}$, i.e., it is proportional to the Bose-Einstein (−) or Fermi-Dirac (+) distribution function, respectively.
[9] P. B. Blakie, A. S. Bradley, M. J. Davis, R. J. Ballagh, and C. W. Gardiner, Advances in Physics **57**, 363 (2008).
[10] Y. Castin, R. Dum, E. Mandonnet, A. Minguzzi, and I. Carusotto, Journal of Modern Optics **47**, 2671 (2000).
[11] To see this, note that the Gibbs ensemble is also correctly sampled if one ignores the term $-\frac{i}{\hbar}\mathcal{L}_\mathbf{C}$ in Eq. (S11). Then the value of $\kappa_{\mathrm{th}}$ only determines the evolution timescale, and is therefore irrelevant to the equilibrium configuration.
[12] C. W. Gardiner, J. R. Anglin, and T. I. A. Fudge, Journal of Physics B: Atomic, Molecular and Optical Physics **35**, 1555 (2002).
[13] C. W. Gardiner and M. J. Davis, Journal of Physics B: Atomic, Molecular and Optical Physics **36**, 4731 (2003).
[14] M. J. Davis, R. J. Ballagh, and K. Burnett, Journal of Physics B: Atomic, Molecular and Optical Physics **34**, 4487 (2001).
[15] M. J. Davis, S. A. Morgan, and K. Burnett, Phys. Rev. Lett. **87**, 160402 (2001).
[16] However, the density is overestimated for non-degenerate gases. Consequently, in a harmonic trap and in the absence of an energy cutoff, the classical field predicts tails of the density that decay as $1/|x|$, leading to an unphysical divergence of the total atom number. Such tails are unimportant for the results that we present here.
[17] S. J. Rooney, P. B. Blakie, and A. S. Bradley, Phys. Rev. E **89**, 013302 (2014).
[18] A. S. Bradley, P. B. Blakie, and C. W. Gardiner, Journal of Physics B: Atomic, Molecular and Optical Physics **38**, 4259 (2005).
[19] P. B. Blakie and M. J. Davis, Phys. Rev. A **72**, 063608 (2005).
[20] A. Sinatra, C. Lobo, and Y. Castin, Journal of Physics B: Atomic, Molecular and Optical Physics **35**, 3599 (2002).

[21] A. Bezett and P. B. Blakie, Phys. Rev. A **79**, 023602 (2009).
[22] T. P. Simula, M. J. Davis, and P. B. Blakie, Phys. Rev. A **77**, 023618 (2008).
[23] I. Bouchoule, M. Arzamasovs, K. V. Kheruntsyan, and D. M. Gangardt, Phys. Rev. A **86**, 033626 (2012).



\end{document}